\documentclass{article}

\PassOptionsToPackage{numbers,compress}{natbib}
\usepackage[preprint]{neurips_2026}

\usepackage[utf8]{inputenc}
\usepackage[T1]{fontenc}
\usepackage{hyperref}
\usepackage{url}
\usepackage{booktabs}
\usepackage{amsfonts}
\usepackage{nicefrac}
\usepackage{microtype}
\usepackage[table]{xcolor}

\usepackage{graphicx}
\usepackage{subcaption}
\usepackage{wrapfig}
\usepackage{amsmath}
\usepackage{amssymb}
\usepackage{mathtools}
\usepackage{amsthm}
\usepackage{algorithm}
\usepackage{algorithmic}
\newcommand{\RETURN}{\STATE \textbf{return} }
\usepackage{listings}
\usepackage{tabularx}
\usepackage{ragged2e}
\lstdefinestyle{icmljson}{
    basicstyle=\ttfamily\scriptsize,
    breaklines=true,
    columns=fullflexible,
    keepspaces=true
}

\usepackage[most]{tcolorbox}
\usepackage{setspace}
\usepackage{fontawesome5}

\definecolor{promptred}{RGB}{220,38,38}
\definecolor{promptblue}{RGB}{32,108,194}
\definecolor{promptpurple}{RGB}{101,16,183}
\definecolor{prompttitle}{RGB}{255,255,255}
\definecolor{promptshadow}{RGB}{170,200,245}
\definecolor{examplepurple}{RGB}{154,108,194}
\definecolor{examplegreen}{RGB}{37,150,190}
\definecolor{mainblue}{RGB}{41, 128, 185}
\definecolor{lightbg}{RGB}{245, 247, 250}
\definecolor{mydarkblue}{rgb}{0,0.08,0.45}
\definecolor{evomainbg}{RGB}{229, 240, 252}
\definecolor{evoablationbg}{RGB}{229, 247, 246}

\usepackage[toc, page, header]{appendix}
\usepackage[capitalize,noabbrev]{cleveref}

\theoremstyle{plain}

\theoremstyle{definition}

\theoremstyle{remark}

\usepackage[textsize=tiny]{todonotes}
\usepackage{multicol}
\usepackage{multirow}
\usepackage{enumitem}
\usepackage{needspace}
\usepackage{arydshln}
\expandafter\let\csname endtabular*\endcsname\endtabular

\title{EvoDesign: Agentic Editable Diagram Creation via Design Expertise Evolution}

\author{%
  \begin{minipage}{\dimexpr\textwidth-2\tabcolsep\relax}
    \centering\normalfont\small
    \textbf{Tianfu Wang$^{1,*,\ddagger}$ \quad Leilei Ding$^{2,*}$ \quad Ziyang Tao$^{2,*}$ \quad Yi Zhan$^{2,*}$ \quad Zhiyuan Ma$^{2}$ \quad Wei Wu$^{2}$}\\[0.3em]
    \textbf{Yuxuan Lei$^{2}$ \quad Junyang Wang$^{2}$ \quad Yizhao Xu$^{3}$ \quad Yin WU$^{1}$ \quad Zhengyu Hu$^{1}$ \quad Hongyuan Zhu$^{2}$}\\[0.3em]
    \textbf{Nicholas Jing Yuan$^{1}$ \quad Yanyong Zhang$^{2}$ \quad Hui Xiong$^{1,4,\dagger}$}\\[0.5em]
    {\footnotesize
      $^{1}$Hong Kong University of Science and Technology (Guangzhou)\\
      $^{2}$University of Science and Technology of China \quad $^{3}$Peking University\\
      $^{4}$Hong Kong University of Science and Technology\\[0.4em]
      $^{*}$Equal contribution \qquad $^{\ddagger}$Project Leader \qquad $^{\dagger}$Corresponding author\\[0.2em]
      $^{\ddagger}$\href{mailto:tianfuwang.cs@gmail.com}{\texttt{tianfuwang.cs@gmail.com}} \quad $^{\dagger}$\href{mailto:xionghui@ust.hk}{\texttt{xionghui@ust.hk}}
    }
  \end{minipage}
}

\begin{document}

\maketitle

\begin{abstract}
High-fidelity diagram creation requires coordinated visual-spatial decisions over semantic topology, visual styling, and geometric layout. Existing methods face a representation gap: pixel-based generation offers limited object-level control, while code-based synthesis provides executable structure at the cost of intuitive manipulation. We introduce EvoDesign, an agentic framework for editable diagram creation through an intermediate canvas schema that is both machine-actionable and directly manipulable by users. EvoDesign first constructs a diagram manifest through coordinated structure, style, and layout agents, and then renders the manifest in a diagnostic verification environment. This environment produces objective defect signals and VLM-based critique, enabling localized refinement of the current diagram while also providing evidence for long-term design expertise evolution. The evolution mechanism distills refinement traces into a three-tier design expertise memory, where candidate strategies are linked to the verification evidence that supports them and are promoted into broader guidelines and principles only through repeated cross-context support. We further define CanvasBench, a canvas-recoverable benchmark and evaluation protocol with content, visual, and cognitive dimensions. Experiments across pixel, code-based, agentic, one-shot schema, and canvas-native methods show that EvoDesign provides the strongest overall balance across content fidelity, visual organization, and cognitive usability. Our code is available at \href{https://github.com/GeminiLight/EvoDesign}{https://github.com/GeminiLight/EvoDesign}.
\end{abstract}

\section{Introduction}

Diagrams such as flowcharts, concept maps, and system architecture graphs serve as ubiquitous visual artifacts for organizing and communicating complex information~\cite{diagram-cognitivescience-1987-diagram}.
By transforming abstract entities and relations into spatial structures, diagrams reduce cognitive load, accelerate comprehension, and support collaborative reasoning.
High-fidelity diagram creation, however, requires more than producing a visually plausible image.
It requires coordinated decisions over semantic topology, visual styling, and geometric layout, where local changes in one layer can alter the communicative meaning of the whole diagram~\cite{diagram-book-2012-diagram-design}.

\begin{figure}
    \centering
    \includegraphics[width=0.86\linewidth]{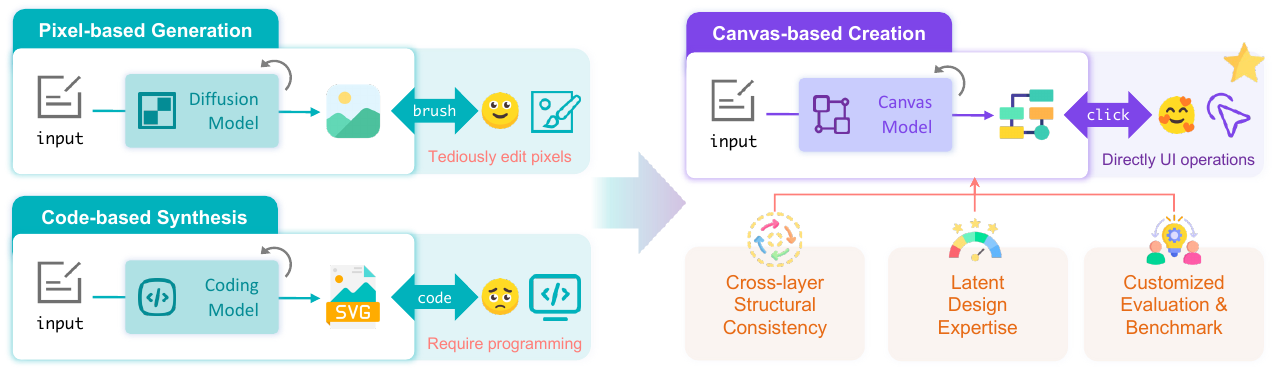}
    \caption{Comparison of diagram generation paradigms. Unlike pixel-based generation (limited control) or code-based synthesis (high barrier), Canvas-based creation unifies AI actionability with human-interpretable UI editing, bridging the representation gap.}
    \label{fig:motivation}
    \vspace{-16pt}
\end{figure}

Existing automated diagramming methods generally rely on either pixel-level generation or executable code synthesis.
Pixel-based generative models can produce visually rich diagram-like renderings, but they provide weak object-level control and may introduce semantic hallucinations, illegible text, or unstable relations~\cite{text2img-cvpr-2023-llm-controlled,tpami-2024-survey-text-to-image,text2img-neurips-2023-layoutgpt}.
Code-based methods, including \LaTeX{}, TikZ, Mermaid, and SVG-oriented synthesis, improve structural executability through symbolic programs~\cite{diagram-cvpr-2025-text2diagram,svg-cvpr-2025-starvector}.
Yet their outputs remain difficult for end users to refine through direct manipulation, since localized changes often require editing low-level syntax rather than moving, regrouping, or restyling diagram objects.
This leaves a representation gap: current formats rarely support autonomous machine generation and intuitive human editing at the same time.

We study editable diagram creation through a canvas-based representation, an intermediate schema composed of object-level entities, relations, styles, and coordinates.
This schema is not only an output format but also an action space for generation, verification, refinement, and user manipulation.
Because the schema can be rendered into a concrete canvas, visual-spatial defects such as overlap, edge crossing, misalignment, and text overflow become observable after execution.
This observation changes the task from one-shot diagram generation into a verifiable design process, where rendered feedback can repair the current artifact and provide reusable experience for future tasks.

Despite this opportunity, realizing canvas-based diagram creation still requires addressing several challenges.
First, editable diagram creation requires cross-layer coordination: semantic structure, visual emphasis, and spatial layout are interdependent and cannot be optimized independently.
Second, reliable feedback is needed for visual-spatial decisions, since many defects only emerge after the canvas is rendered and inspected.
Third, high-quality diagrams depend on latent, domain-dependent design expertise, such as layout conventions, density control, and visual hierarchy, which should be accumulated and retrieved rather than repeatedly rediscovered from scratch.

To address these challenges, we introduce EvoDesign, an agentic framework for editable diagram creation via design expertise evolution.
EvoDesign constructs a canvas manifest through a coordinated multi-agent pipeline for semantic structure, visual styling, and spatial layout.
It then executes the manifest in a diagnostic verification environment that combines objective programmatic checks with VLM-based critique.
The verification output drives a localized refiner to correct the current diagram and provides structured evidence for design expertise evolution.
Through this mechanism, EvoDesign distills refinement traces into a three-tier memory of sample strategies, domain guidelines, and general principles, where each abstraction preserves the diagnostic evidence that supports it and can be retrieved by agents in later tasks.

Our contributions are summarized as follows:
\begin{itemize}[leftmargin=*, nosep, topsep=2pt]
    \item We formulate editable canvas creation as a diagram generation paradigm that bridges pixel-level synthesis and code-level execution through a machine-actionable, user-editable schema.
    \item We propose verification-guided design expertise evolution, which converts diagnostic refinement traces into a maintained hierarchy of reusable design knowledge.
    \item We define CanvasBench, a canvas-recoverable benchmark and evaluation protocol with content, visual, and cognitive dimensions for comparing diagram generation methods.
\end{itemize}

\section{Related Work}

\textbf{Automated Diagram Generation.}
Existing approaches to automated diagram generation mainly rely on two underlying data representations: pixel arrays and symbolic code.
Advanced by text-to-image models, pixel-based approaches~\cite{pixel-wan2025wan,pixel-labs2025flux,pixel-xie2024sana} excel in generating visually diverse renderings. However, these models suffer from stochasticity, frequently producing semantic hallucinations (e.g., illegible text or chaotic relationships) and static, rasterized artifacts. This lack of object-level separability renders post-generation editing impossible without complex inpainting.
To ensure structural integrity, research has shifted toward generating executable code. Recent works~\cite{carlier2020deepsvg,rodriguez2025starvector} focus on synthesizing low-level vector primitives in SVG code to reconstruct intricate geometric details with resolution independence. Alternatively, other works use programmatic languages, such as \LaTeX-based TikZ~\cite{diagram-cvpr-2025-text2diagram,diagram-ase-2025-tikz} or domain-specific languages like Mermaid~\cite{deka2025flowchart2mermaid}, which prioritize semantic constraints and structural validity.
However, this paradigm introduces a high usability barrier. The resulting representations are brittle for end-users, as localized adjustments require manipulating low-level syntax rather than intuitive visual handles.
There lacks a unified representation that is both machine-actionable for generation and human-interpretable for manipulation. EvoDesign addresses this by adopting a canvas-based schema, effectively bridging the gap between the generative precision of symbolic code and the intuitive flexibility of modern design interfaces.

\textbf{Agentic Media Creation.}
Multimodal agents now orchestrate complex visual media such as posters, slides, and videos~\cite{agent-arxiv-2025-vllm-agent-survey,agenticcreation-arxiv-2026-survey,llm-arxiv-2026-kimi-k3}.
For paper posters, Paper2Poster~\cite{agenticcreation-neurips-2025-paper2poster} and PosterGen~\cite{agenticcreation-arxiv-2025-postergen} use multi-agent systems~\cite{wu2024autogen} to condense scientific content into a single visual plane.
For multi-page narratives, PPTAgent~\cite{agenticcreation-arxiv-2025-pptagent} and SlideTailor~\cite{agenticcreation-aaai-2026-slidetailor} automate slide creation by modeling logical flow and structure.
For dynamic media, Code2Video~\cite{creation-arxiv-code2video}, Paper2Video~\cite{agenticcreation-arxiv-2025-paper2video}, and PresentAgent~\cite{agenticcreation-emnlp-demo-2025-presentagent} coordinate scripts, visuals, transitions, and voiceovers.
While these frameworks treat media as macro-containers for asset arrangement, diagrams are compact, structure-dense kernels~\cite{diagram-cognitivescience-1987-diagram} that translate abstract semantics into concrete topology within a confined space.
Human--AI coding collaboration similarly emphasizes inspectable, revisable shared state~\cite{collaboration-arxiv-2026-consensus}.
EvoDesign studies agentic creation for this topology-centric medium with an automated expertise evolution perspective.

\textbf{Agent Memory Evolution.}
Memory enables agents to move from stateless inference toward continual adaptation~\cite{memory-arxiv-memory-survey}.
Prior work studies both memory architecture and active evolution~\cite{evolution-arxiv-2026-hase,evolution-arxiv-2026-meta-team,evolution-arxiv-2026-tact}: MemoryOS~\cite{memory-arxiv-2025-memoryos} uses a hierarchical framework, A-MEM~\cite{memory-arxiv-2025-amem} organizes interconnected knowledge, and MIRIX~\cite{memory-arxiv-2025-mirix} coordinates memory updates in multi-agent settings.
User modeling supports behavior-aware prediction~\cite{profiling-kdd-2024-comet}, personalized simulation~\cite{personalization-arxiv-2026-humanllm}, population-aligned persona generation~\cite{simulation-arxiv-2025-persona}, and personalized tutoring and practice~\cite{wang2025llm,tutoring-arxiv-2026-socialcoach}.
AlphaEvolve~\cite{memory-arxiv-2025-alphaevolve} evolves code libraries, EvoMem~\cite{memory-arxiv-2025-evomem} uses dual-evolving memory for planning, ReasoningBank~\cite{memory-arxiv-2025-reasoningbank} distills reasoning patterns, and ACE~\cite{memory-arxiv-2025-ace} treats context as an evolving playbook.
High-fidelity diagramming requires implicit expertise over multi-objective visual-spatial trade-offs.
EvoDesign therefore distills role-specific design expertise into a hierarchical memory to improve multi-agent synergy and global consistency.

\begin{figure*}
    \centering
    \includegraphics[width=0.98\linewidth]{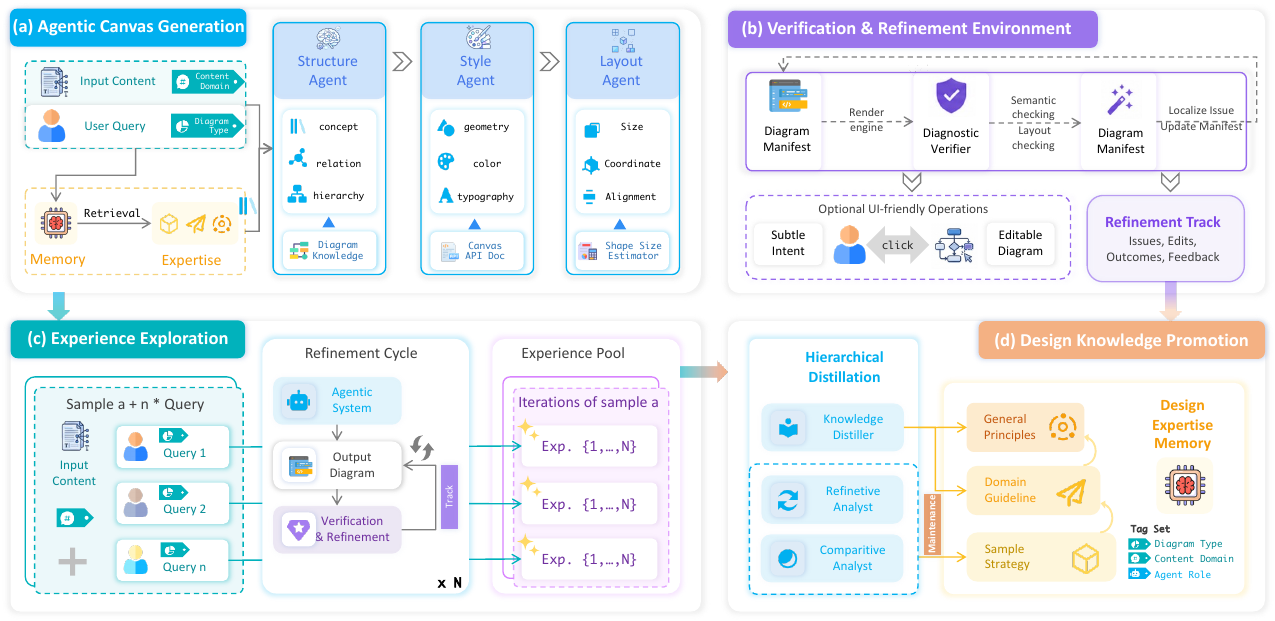}
    \vspace{-4pt}
    \caption{Overview of the EvoDesign framework. \textit{(a) Agentic Canvas Generation}: EvoDesign transforms input content and user queries into a diagram manifest through structure, style, and layout agents, with retrieved expertise and canvas API knowledge grounding the generation process. \textit{(b) Verification and Refinement Environment}: the manifest is rendered into an editable diagram, checked by a diagnostic verifier, and locally repaired through issue localization and manifest updates, while optional UI-friendly operations and refinement tracks record issues, edits, outcomes, and feedback. \textit{(c) Experience Exploration}: multiple query variants for the same content are sampled and repeatedly passed through agentic generation, verification, and refinement to collect comparable experience trajectories. \textit{(d) Design Knowledge Promotion}: refinement tracks are distilled and analyzed into sample strategies, domain guidelines, and general principles, which are maintained with tag sets and stored in the design expertise memory.}
    \vspace{-10pt}
    \label{fig:framework}
\end{figure*}

\section{Task Formulation}
\label{sec:task}

Let $\mathcal{C}$ denote the space of semantic contents providing the informational context, and $\mathcal{R}$ represent optional user requirements specifying preferences for content, style, or layout. We define the diagram generation task as a mapping from $c$ and $r$ to an object-level editable representation $D$, i.e., $\mathcal{F}_{\pi}: I = (c, r) \rightarrow D$, where $\pi$ is a design policy, $c \in \mathcal{C}$ and $r \in \mathcal{R} \cup \{\emptyset\}$.
Formally, the diagram representation $D$ is defined as $D = (G, S, L)$ with the following components:
\begin{itemize}[leftmargin=*, nosep, topsep=0pt]
    \item \textit{Semantic Graph $G = (\mathcal{V}, \mathcal{E})$}: A structural abstraction where nodes $\mathcal{V}$ and edges $\mathcal{E}$ capture the logical entities and relational dependencies extracted from the content $c$.
    \item \textit{Style Schema $S$}: A set of visual encodings that assigns attributes (e.g., color, shape, fill) to each element of $G$ based on design principles and user preferences $r$.
    \item \textit{Layout Configuration $L$}: A geometric arrangement on a canvas defining shape dimensions, spatial coordinates, and routing functions to consider perceptual clarity and $r$.
\end{itemize}
By executing such a representation within a canvas-based rendering environment (e.g., \texttt{excalidraw}), we obtain a visual artifact $\hat{D}$ that is inherently interactive and UI-friendly.
The representation is therefore not only an output format, but also an action space for automated generation, diagnostic verification, localized refinement, and direct user manipulation.
This allows users to perform post-generation refinements via familiar UI operations while enabling the system to inspect and modify object-level diagram states programmatically.
The objective is to synthesize a feasible and effective $D^*$ that balances semantic faithfulness, visual consistency, and spatial resolvability, while satisfying the user requirements $r$ if provided.

\section{Agentic EvoDesign}

EvoDesign operationalizes editable diagram creation through a design policy $\pi=\{\mathcal{A},\mathcal{M}\}$, where $\mathcal{A}$ is a coordinated set of generation and refinement agents and $\mathcal{M}$ is an evolving design expertise memory.
Given an input $I=(c,r)$, the system first produces a canvas manifest $D=(G,S,L)$ through multi-agent generation, then renders and verifies the manifest to obtain structured diagnostic feedback.
This feedback has two roles: it guides localized repair of the current diagram and provides evidence used to distill reusable design expertise.
The overall process therefore links short-term diagram correction with long-term expertise accumulation.
Algorithm~\ref{alg:generation} and Algorithm~\ref{alg:evolution} in Appendix~\ref{app:algorithm} provide the corresponding pseudocode for the generation loop and the expertise evolution cycle.

\subsection{Multi-Agent Canvas Generation}

The initial manifest is produced by a coordinated generation backbone with three specialized agents.
Each agent follows a spec-to-instance pattern: it first forms a high-level design specification conditioned on the input and retrieved expertise, and then instantiates concrete schema fields that are valid under the canvas renderer.
This pattern separates abstract design intent from low-level canvas constraints, reducing hallucinated properties while preserving object-level editability.

\textbf{Semantic Structure Agent.}
The structure agent maps the source content into a semantic graph $G=(\mathcal{V},\mathcal{E})$.
It first determines the diagram type, narrative scope, hierarchy, and relational motif needed to express the user intent.
It then grounds entities into canonical nodes, removes peripheral noise, nests related concepts into groups, and constructs salient relations.
This entity-first, relation-last process provides a stable symbolic topology for downstream styling and layout.

\textbf{Visual Style Agent.}
The style agent assigns visual encodings $S$ to nodes, groups, and edges in the semantic graph.
It first builds a design specification covering palette, visual hierarchy, shape vocabulary, and emphasis strategy.
The specification is then translated into renderer-supported attributes such as color, fill, stroke, and shape type.
By grounding abstract style intent into a finite property space, the agent maintains visual consistency and avoids unsupported canvas states.

\textbf{Spatial Layout Agent.}
The layout agent computes the geometric configuration $L$ for the styled graph.
It uses tool-augmented size estimation to account for text wrapping, group padding, and element footprints before assigning coordinates.
The agent then performs recursive placement, resolving global arrangement for top-level modules and local coordinates within nested groups.
This produces an executable layout manifest with positions, dimensions, ordering, and edge routing information.

\subsection{Diagnostic Verification and Localized Refinement}
\label{sec:method:refinement}

Once a manifest is rendered into a canvas artifact $\hat{D}$, EvoDesign evaluates it in a \textit{Diagnostic Verification Environment}.
The environment produces two complementary forms of feedback.
Objective checks identify programmatic visual-spatial defects, including node overlap, edge crossing, alignment error, text overflow, and canvas utilization.
VLM-based critique provides higher-level assessments of semantic completeness, visual hierarchy, and aesthetic coherence that are difficult to express as deterministic rules.

The verification output is represented as a prioritized finding list $\mathcal{F}=\{f_i\}_{i=1}^{n}$.
Each finding records the defect category, severity, affected elements, and a natural-language description.
Both the objective score $S_{\mathrm{obj}}$ and VLM critique score $S_{\mathrm{crit}}$ use a 0--10 scale.
Objective findings incur a severity-weighted penalty $P=5n_{\mathrm{critical}}+3n_{\mathrm{major}}+n_{\mathrm{minor}}+0.2n_{\mathrm{info}}$, giving $S_{\mathrm{obj}}=10[1-\min(P,50)/50]$; fewer defects yield a higher score.
The critique score follows a fixed rubric for semantic and visual quality.
In the reported experiments, verification combines these signals and measures refinement improvement as
\begin{equation}
\begin{aligned}
\operatorname{score}(\hat{D}) &= 0.6S_{\mathrm{obj}}(\hat{D})+0.4S_{\mathrm{crit}}(\hat{D}),\\
\Delta_{\text{score}}^{(t)} &= \operatorname{score}(\hat{D}^{(t+1)})-\operatorname{score}(\hat{D}^{(t)}).
\end{aligned}
\label{eq:verification-score}
\end{equation}
This score guides local refinement and strategy selection.

The \textit{Localized Refiner Agent} consumes $\mathcal{F}$ and applies targeted edits to the manifest instead of regenerating the whole diagram.
Its action space includes single-element operations, such as moving a node or changing a style attribute, and cross-element operations, such as aligning a group or rerouting an edge.
After each repair, the diagram is rendered and verified again.
This loop improves the current artifact while preserving the semantic structure established by the generation stage.

The finding list and $\Delta_{\text{score}}$ also support design expertise evolution by linking corrective actions to their observed effects.
This generation and verification loop is summarized in Algorithm~\ref{alg:generation}.

\subsection{Design Expertise Evolution}
\label{sec:method:evolution}

Design quality depends on latent visual-spatial heuristics that vary across diagram types, domains, and agent roles.
EvoDesign turns such heuristics into reusable expertise through four coupled operations: storing refinement evidence in a hierarchical memory, exploring experience through verification-guided refinement, promoting supported strategies across contexts, and maintaining the resulting memory through usage-aware updates.
Algorithm~\ref{alg:evolution} formalizes this evolution cycle.

\textbf{Hierarchical Expertise Memory.}
EvoDesign maintains a design expertise memory $\mathcal{M}=\{\mathcal{K}^{s},\mathcal{K}^{g},\mathcal{K}^{p}\}$ with three levels of abstraction.
The bottom tier $\mathcal{K}^{s}$ stores \textit{sample strategies}: local, trace-grounded rules that describe how a concrete refinement action resolved a specific diagram failure.
The middle tier $\mathcal{K}^{g}$ stores \textit{domain guidelines}: strategies that receive repeated support across related domains or task contexts.
The top tier $\mathcal{K}^{p}$ stores \textit{general principles}: higher-level design rules that remain useful across diagram types, such as density control, modular separation, or role-consistent visual hierarchy.
Each memory item records its tier, responsible agent role, diagram type, domain context, precondition, corrective action, expected effect, evidence identifiers, confidence, and usage statistics.
This schema makes each abstraction auditable: a principle can be traced back to the guidelines and sample strategies that supported it, while each local strategy remains tied to its original refinement evidence.

\textbf{Verification-Guided Experience Exploration.}
Experience is collected from the same rendered feedback loop used for localized refinement.
For a given diagram, vertical exploration follows the refinement trajectory over multiple steps.
At each step, the system records an evidence unit containing the pre-repair findings, applied edits, post-repair findings, and $\Delta_{\text{score}}$.
This step-level record estimates whether a corrective action reduced a concrete visual-spatial failure.
To test whether an action remains useful beyond a single prompt, horizontal exploration creates nearby task variants, such as alternative user requirements or contextual constraints for the same content, and runs the generation-refinement loop again.
Together, vertical and horizontal evidence turn individual repairs into comparable experience for later expertise evolution.

\textbf{Evidence-Guided Expertise Promotion.}
The evolution process converts refinement experience into increasingly reusable expertise while preserving provenance.
At the base tier, EvoDesign extracts candidate strategies from verified refinement traces.
Each candidate strategy is linked to the refinement steps, edit operations, and resolved finding categories that support it.
Its evidence score is the mean $\Delta_{\text{score}}$ over its supporting steps, and the candidate enters $\mathcal{K}^{s}$ only when this mean exceeds $\tau=0.1$.
The mixture weights and $\tau$ were fixed on a held-out split before evaluation samples were inspected.
Related strategies are then grouped by agent role, diagram context, and action template.
Promotion into $\mathcal{K}^{g}$ requires at least five supporting traces spanning three domains, with links to the source strategies.
Promotion into $\mathcal{K}^{p}$ retains this evidence floor and additionally requires support across at least two diagram types.
Promotion therefore expands the scope and confidence of a rule without discarding the evidence needed for later inspection.

\textbf{Active Knowledge Maintenance.}
The memory is periodically maintained through scoring, pruning, merging, promotion, and demotion.
Confidence combines creation-time evidence from refinement traces with deployment-time evidence from later retrieval outcomes, support coverage, and redundancy checks.
Low-utility items are pruned or demoted, semantically redundant items are merged, and high-support items are promoted when later evidence shows broader transferability.
During generation, agents retrieve expertise by filtering on agent role, diagram type, and domain, then ranking candidates by semantic relevance and confidence.
This retrieval closes the loop: maintained expertise guides future multi-agent generation, while new verified traces update the memory in subsequent evolution rounds.

\section{The CanvasBench}
Existing diagrammatic benchmarks often contain high-entropy artifacts (e.g., photorealistic textures, complex data visualizations) that are incompatible with vector-based canvas schemas.
To bridge this gap, we introduce CanvasBench, a dataset filtered for \textit{diagram editability}, requiring structures that can be represented as editable objects, accompanied by a multi-dimensional evaluation framework spanning content, visual, and cognitive perspectives.

\begin{figure*}
    \centering
    \includegraphics[width=0.96\linewidth]{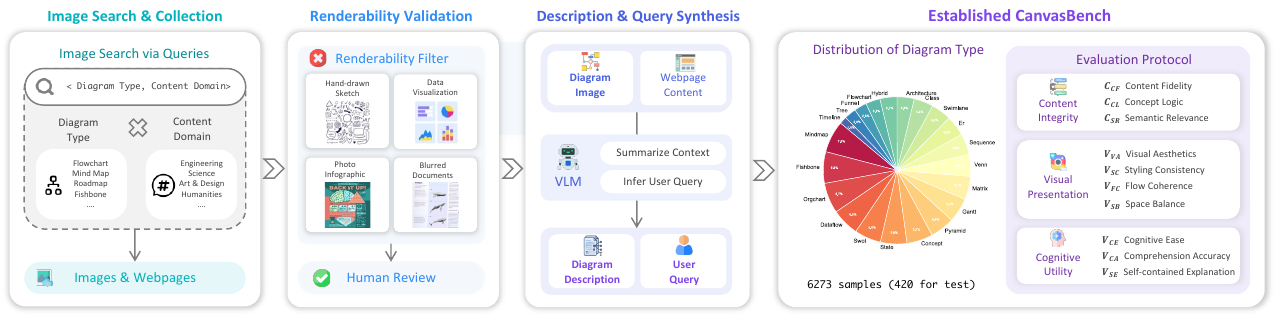}
    \caption{The dataset construction pipeline and the overview of CanvasBench.}
    \label{fig:benchmark}
    \vspace{-8pt}
\end{figure*}

\subsection{Data Collection Pipeline}
\label{sec:benchmark:collection}
As illustrated in Figure~\ref{fig:benchmark}, we implement a three-stage pipeline to synthesize diagram-instruction pairs. By intersecting 21 curated diagram types with 30 vertical domains from the MMMU benchmark, we retrieve candidate images and use Qwen3-VL-Max to filter out high-entropy artifacts (e.g., sketches and photo infographics). Two annotators independently review candidate images under written guidelines, with Cohen's $\kappa=0.74$. URL-level filtering excludes repeated source pages. Image pairs with high embedding similarity are reviewed by a VLM, and low-confidence cases are escalated to human review. No additional near-duplicate images were identified by this review. Finally, a VLM reconstructs dense descriptions and canonical user queries from the validated images and their webpage contexts, producing (image, query, content) triplets. The synthesized descriptions are manually checked for sufficient drawing information, and the description prompts are refined using this feedback. See Appendix~\ref{app:benchmark:collection} for the search taxonomy and synthesis procedure.

CanvasBench focuses on structured, editable diagrams with reference-derived descriptions and queries.

\subsection{Dataset Statistics}
Finally, CanvasBench contains 6,273 validated samples across 21 diagram types, consisting of 5,853 samples for training and design expertise evolution and a balanced held-out test set of 420 samples (20 per diagram type). Each sample ensures object-level canvas editability on diverse, real-world visual topologies. As illustrated in Figure~\ref{fig:benchmark}, the training/evolution split preserves broad coverage over diagram categories, while the test split controls for type frequency to support comparable evaluation.
The samples preserve heterogeneous layout styles from human-authored diagrams. Further analysis and examples appear in Appendix~\ref{app:benchmark:analysis} and Appendix~\ref{app:benchmark:examples}.

\subsection{Evaluation Protocol}
\label{sec:benchmark:evaluation}
Rigorous diagram assessment extends beyond semantic correctness and visual presentation to ensure \textit{cognitive interpretability}.
To capture this, we propose a multi-dimensional evaluation framework structured along three primary axes. Detailed definitions are provided in Appendix~\ref{app:benchmark:metrics}. 
VLM-based judges assign scores on a 5-point Likert scale.

\begin{itemize}[leftmargin=*, nosep, topsep=0pt]
    \item \textit{Content Integrity Dimension ($\mathcal{C}$)} evaluates semantic alignment between the source text and the generated schema via \textit{Content Fidelity} ($\text{C}_{\textit{CF}}$), {Concept Logic} ($\text{C}_{\textit{CL}}$), and {Semantic Relevance} ($\text{C}_{\textit{SR}}$).
    \item \textit{Visual Presentation Dimension ($\mathcal{V}$)} assesses the structural and aesthetic validity of the canvas style and layout through  {Visual Aesthetics} ($\text{V}_{\textit{VA}}$), {Styling Consistency} ($\text{V}_{\textit{SC}}$), {Flow Coherence} ($\text{V}_{\textit{FC}}$), and {Space Balance} ($\text{V}_{\textit{SB}}$).
    \item \textit{Cognitive Utility Dimension ($\mathcal{G}$)} quantifies the cognitive efficiency and interpretability for information translation using {Cognitive Ease} ($\text{G}_{\textit{CE}}$), {Comprehension Accuracy} ($\text{G}_{\textit{CA}}$), and {Self-contained Explanation} ($\text{G}_{\textit{SE}}$).
\end{itemize}

Cognitive Ease estimates the effort required to read the visual structure, while Comprehension Accuracy uses VLM-based question answering to check whether factual and structural information can be recovered solely from the diagram.
Self-contained Explanation evaluates whether the diagram conveys its narrative without access to the source text.

\section{Experiments}
\label{sec:evaluation}
In this section, we evaluate EvoDesign against representative baselines on CanvasBench.

\begin{table*}[t]
\centering
\caption{Main results on CanvasBench across content, visual, and cognitive dimensions.}
\label{tab:main_results}
\scriptsize
\setlength{\tabcolsep}{2.2pt}
\resizebox{0.99\textwidth}{!}{
\begin{tabular}{lll ccc cccc ccc}
\toprule
\textbf{Category} & \textbf{Method} & \textbf{Model} & \multicolumn{3}{c}{\textbf{Content}} & \multicolumn{4}{c}{\textbf{Visual}} & \multicolumn{3}{c}{\textbf{Cognitive}} \\
\cmidrule(lr){4-6} \cmidrule(lr){7-10} \cmidrule(lr){11-13}
& & & $\text{C}_{\textit{CF}}$ & $\text{C}_{\textit{CL}}$ & $\text{C}_{\textit{SR}}$ & $\text{V}_{\textit{VA}}$ & $\text{V}_{\textit{SC}}$ & $\text{V}_{\textit{FC}}$ & $\text{V}_{\textit{SB}}$ & $\text{G}_{\textit{CE}}$ & $\text{G}_{\textit{CA}}$ & $\text{G}_{\textit{SE}}$ \\
\midrule
Diffusion & GPT-4o-Image & -- & 2.111 & 2.349 & 2.444 & 3.810 & 3.635 & 3.651 & 3.571 & 4.048 & 2.270 & 2.921 \\
Diffusion & Gemini-2.5-Flash-Image & -- & 1.857 & 2.127 & 2.111 & 3.206 & 2.889 & 2.857 & 3.317 & 2.889 & 2.000 & 2.476 \\
Diffusion & Flux.2 flex & -- & 1.825 & 2.000 & 2.063 & 3.254 & 2.952 & 2.952 & 3.254 & 2.937 & 1.857 & 2.397 \\
\hdashline
Code-based & \LaTeX{} TikZ & GPT-5.5 & 3.404 & 3.421 & 3.579 & 2.667 & 2.579 & 3.246 & 2.930 & 3.772 & 3.035 & 2.789 \\
Code-based & \LaTeX{} TikZ & Gemini-3.1-Pro & 4.021 & 3.932 & 4.018 & 3.184 & 3.438 & 3.512 & 3.076 & 3.704 & 3.352 & 2.604 \\
Code-based & \LaTeX{} TikZ & DeepSeek-V4-Pro & 3.884 & 3.792 & 3.946 & 3.012 & 3.214 & 3.281 & 2.846 & 3.486 & 3.186 & 2.482 \\
Code-based & \LaTeX{} TikZ & DeepSeek-V4-Flash & 3.526 & 3.412 & 3.561 & 2.702 & 2.954 & 2.986 & 2.512 & 3.108 & 2.714 & 2.206 \\
\hdashline
Agentic & DiagramAgent & GPT-5.5 & 2.000 & 2.476 & 1.921 & 2.778 & 2.825 & 2.683 & 2.540 & 2.349 & 2.508 & 2.492 \\
Agentic & DiagramAgent & Gemini-3.1-Pro & 2.842 & 3.102 & 2.784 & 3.042 & 3.168 & 3.044 & 2.861 & 2.934 & 2.886 & 2.612 \\
Agentic & DiagramAgent & DeepSeek-V4-Pro & 2.716 & 2.948 & 2.663 & 2.884 & 3.031 & 2.872 & 2.702 & 2.781 & 2.741 & 2.486 \\
Agentic & DiagramAgent & DeepSeek-V4-Flash & 2.346 & 2.624 & 2.281 & 2.641 & 2.812 & 2.563 & 2.418 & 2.512 & 2.438 & 2.214 \\
Agentic & SciSketch & GPT-5.5 & 4.048 & 3.794 & 3.587 & 2.984 & 3.063 & 2.921 & 2.841 & 3.159 & 3.540 & 3.222 \\
Agentic & SciSketch & Gemini-3.1-Pro & 4.108 & 3.906 & 3.784 & 3.214 & 3.302 & 3.182 & 3.014 & 3.426 & 3.612 & 3.264 \\
Agentic & SciSketch & DeepSeek-V4-Pro & 4.006 & 3.832 & 3.712 & 3.072 & 3.188 & 3.064 & 2.902 & 3.281 & 3.496 & 3.146 \\
Agentic & SciSketch & DeepSeek-V4-Flash & 3.724 & 3.514 & 3.408 & 2.812 & 2.986 & 2.824 & 2.657 & 3.014 & 3.146 & 2.884 \\
\midrule
Schema & Prompting & GPT-5.5 & 3.884 & 3.566 & 3.612 & 2.463 & 2.891 & 2.512 & 2.064 & 2.806 & 2.965 & 2.486 \\
Schema & Prompting & Gemini-3.1-Pro & 3.812 & 3.741 & 3.829 & 2.752 & 3.116 & 2.866 & 2.351 & 3.168 & 3.034 & 2.604 \\
Schema & Prompting & DeepSeek-V4-Pro & 3.738 & 3.664 & 3.758 & 2.557 & 2.998 & 2.674 & 2.136 & 2.914 & 2.898 & 2.376 \\
Schema & Prompting & DeepSeek-V4-Flash & 3.438 & 3.274 & 3.371 & 2.276 & 2.742 & 2.412 & 1.886 & 2.586 & 2.482 & 2.008 \\
\rowcolor{evomainbg}Canvas-native & \textbf{EvoDesign} & GPT-5.5 & 4.118 & 4.052 & 4.126 & 3.612 & 3.786 & 3.704 & 3.536 & 3.912 & 3.512 & 3.086 \\
\rowcolor{evomainbg}Canvas-native & \textbf{EvoDesign} & Gemini-3.1-Pro & 4.062 & 4.018 & 4.097 & 3.526 & 3.714 & 3.681 & 3.472 & 3.861 & 3.486 & 3.054 \\
\rowcolor{evomainbg}Canvas-native & \textbf{EvoDesign} & DeepSeek-V4-Pro & 3.936 & 3.887 & 4.012 & 3.317 & 3.562 & 3.421 & 3.213 & 3.674 & 3.342 & 2.891 \\
\rowcolor{evomainbg}Canvas-native & \textbf{EvoDesign} & DeepSeek-V4-Flash & 3.684 & 3.602 & 3.758 & 3.038 & 3.326 & 3.082 & 2.842 & 3.386 & 3.012 & 2.548 \\
\bottomrule
\end{tabular}
}
\vspace{-10pt}
\end{table*}

\subsection{Experimental Settings}

\textbf{Evaluation Metrics}.
We adopt the \textit{CanvasBench} evaluation protocol defined in Section~\ref{sec:benchmark:evaluation}, where VLM-based judges assign scores on a 5-point Likert scale across content, visual, and cognitive dimensions.
For the cross-model rows, we report the mean score of each metric over the balanced evaluation subset.

\textbf{Comparative Baselines}.
We compare against pixel-based generators, \LaTeX{} TikZ code synthesis, one-shot canvas-schema generation, and agentic diagram systems.
For the non-pixel baselines, we evaluate the same method under GPT-5.5, Gemini-3.1-Pro, DeepSeek-V4-Pro, and DeepSeek-V4-Flash whenever the method can use a backbone VLM.
All methods are evaluated on the same balanced CanvasBench subset, with task prompts normalized to expose the same source content and diagram intent.
EvoDesign uses the \texttt{excalidraw} canvas schema as the executable representation, with the agent workflow orchestrated by \textsc{LangGraph}.
Unless otherwise stated, generation uses temperature 0.7, top-$k$ retrieval from the design expertise memory, and one localized refinement pass after diagnostic feedback.
Additional implementation details, data splits, and baseline settings are provided in Appendix~\ref{app:experiment}.

\subsection{Main Results}

EvoDesign exhibits a balanced performance profile across content, visual, and cognitive axes.
As shown in Table~\ref{tab:main_results}, code-based and agentic baselines can be strong on selected content metrics, especially \LaTeX{} TikZ and SciSketch under stronger backbones.
However, they remain less consistent on canvas-native visual organization and cognitive usability.
By contrast, EvoDesign achieves the strongest overall balance across the four backbone models.
Its most pronounced gains appear on styling fitness, space balance, cognitive ease, and self-contained explanation, matching the goal of editable diagram creation rather than static image fidelity alone.

The cross-model rows also show that the improvement is not tied to a single VLM.
Under GPT-5.5, EvoDesign improves over one-shot prompting by $+1.472$ on Space Balance and $+1.106$ on Cognitive Ease.
The same pattern holds for Gemini-3.1-Pro, DeepSeek-V4-Pro, and DeepSeek-V4-Flash, where EvoDesign remains consistently ahead of prompting and competitive external baselines on the visual and cognitive dimensions.

\subsection{Ablation Study}

We conduct an ablation study in Table~\ref{tab:ablation_results} to isolate the multi-agent decomposition, design expertise memory, and localized refiner across backbone models.
The w/o Multiagent setting replaces role-specialized structure, style, and layout planning with a less decomposed generation process; w/o Memory disables retrieved design expertise; and w/o Refiner removes the localized repair step after diagnostic feedback.
Across all backbones, the full system consistently outperforms these variants.
The largest drops appear when removing multi-agent decomposition, especially on Space Balance, while memory and refinement provide complementary gains on reusable design priors and local visual repair.

\begin{table*}[t]
\centering
\caption{Ablation results across backbone models.}
\label{tab:ablation_results}
\scriptsize
\setlength{\tabcolsep}{2.4pt}
\resizebox{0.99\textwidth}{!}{
\begin{tabular}{ll ccc cccc ccc}
\toprule
\textbf{Model} & \textbf{Variant} & \multicolumn{3}{c}{\textbf{Content}} & \multicolumn{4}{c}{\textbf{Visual}} & \multicolumn{3}{c}{\textbf{Cognitive}} \\
\cmidrule(lr){3-5} \cmidrule(lr){6-9} \cmidrule(lr){10-12}
& & $\text{C}_{\textit{CF}}$ & $\text{C}_{\textit{CL}}$ & $\text{C}_{\textit{SR}}$ & $\text{V}_{\textit{VA}}$ & $\text{V}_{\textit{SC}}$ & $\text{V}_{\textit{FC}}$ & $\text{V}_{\textit{SB}}$ & $\text{G}_{\textit{CE}}$ & $\text{G}_{\textit{CA}}$ & $\text{G}_{\textit{SE}}$ \\
\midrule
\rowcolor{evomainbg}GPT-5.5 & \textbf{EvoDesign} & 4.118 & 4.052 & 4.126 & 3.612 & 3.786 & 3.704 & 3.536 & 3.912 & 3.512 & 3.086 \\
\rowcolor{evoablationbg}GPT-5.5 & \textit{w/o Multiagent} & 3.946 & 3.901 & 3.982 & 3.184 & 3.426 & 3.298 & 2.824 & 3.512 & 3.246 & 2.812 \\
\rowcolor{evoablationbg}GPT-5.5 & \textit{w/o Memory} & 4.036 & 3.972 & 4.041 & 3.428 & 3.566 & 3.482 & 3.178 & 3.684 & 3.402 & 2.938 \\
\rowcolor{evoablationbg}GPT-5.5 & \textit{w/o Refiner} & 4.074 & 4.006 & 4.083 & 3.361 & 3.498 & 3.548 & 3.096 & 3.642 & 3.431 & 2.951 \\
\hdashline
\rowcolor{evomainbg}Gemini-3.1-Pro & \textbf{EvoDesign} & 4.062 & 4.018 & 4.097 & 3.526 & 3.714 & 3.681 & 3.472 & 3.861 & 3.486 & 3.054 \\
\rowcolor{evoablationbg}Gemini-3.1-Pro & \textit{w/o Multiagent} & 3.912 & 3.887 & 3.962 & 3.126 & 3.338 & 3.245 & 2.752 & 3.476 & 3.214 & 2.781 \\
\rowcolor{evoablationbg}Gemini-3.1-Pro & \textit{w/o Memory} & 3.976 & 3.941 & 4.006 & 3.348 & 3.472 & 3.441 & 3.136 & 3.621 & 3.367 & 2.872 \\
\rowcolor{evoablationbg}Gemini-3.1-Pro & \textit{w/o Refiner} & 4.018 & 3.972 & 4.041 & 3.286 & 3.421 & 3.512 & 3.048 & 3.584 & 3.402 & 2.884 \\
\hdashline
\rowcolor{evomainbg}DeepSeek-V4-Pro & \textbf{EvoDesign} & 3.936 & 3.887 & 4.012 & 3.317 & 3.562 & 3.421 & 3.213 & 3.674 & 3.342 & 2.891 \\
\rowcolor{evoablationbg}DeepSeek-V4-Pro & \textit{w/o Multiagent} & 3.811 & 3.786 & 3.892 & 2.952 & 3.171 & 3.018 & 2.541 & 3.247 & 3.056 & 2.612 \\
\rowcolor{evoablationbg}DeepSeek-V4-Pro & \textit{w/o Memory} & 3.872 & 3.831 & 3.936 & 3.156 & 3.294 & 3.217 & 2.941 & 3.426 & 3.204 & 2.718 \\
\rowcolor{evoablationbg}DeepSeek-V4-Pro & \textit{w/o Refiner} & 3.918 & 3.846 & 3.954 & 3.091 & 3.238 & 3.263 & 2.864 & 3.392 & 3.241 & 2.731 \\
\hdashline
\rowcolor{evomainbg}DeepSeek-V4-Flash & \textbf{EvoDesign} & 3.684 & 3.602 & 3.758 & 3.038 & 3.326 & 3.082 & 2.842 & 3.386 & 3.012 & 2.548 \\
\rowcolor{evoablationbg}DeepSeek-V4-Flash & \textit{w/o Multiagent} & 3.556 & 3.498 & 3.627 & 2.674 & 2.952 & 2.684 & 2.236 & 3.027 & 2.718 & 2.238 \\
\rowcolor{evoablationbg}DeepSeek-V4-Flash & \textit{w/o Memory} & 3.621 & 3.546 & 3.689 & 2.862 & 3.084 & 2.904 & 2.612 & 3.164 & 2.884 & 2.371 \\
\rowcolor{evoablationbg}DeepSeek-V4-Flash & \textit{w/o Refiner} & 3.672 & 3.562 & 3.701 & 2.791 & 3.036 & 2.942 & 2.514 & 3.132 & 2.912 & 2.386 \\
\bottomrule
\end{tabular}
}
\vspace{-10pt}
\end{table*}

\subsection{Knowledge Evolution Dynamics}

\begin{figure*}[t]
    \centering
    \vspace{2pt}
    \includegraphics[width=0.82\linewidth]{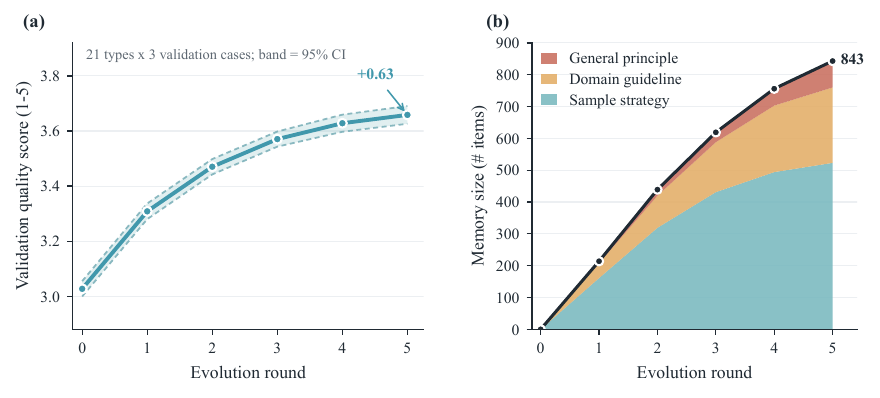}
    \caption{Evolution dynamics of the design expertise memory. Left: validation quality improves as refinement traces accumulate across evolution rounds. Right: the memory grows from sample-level strategies into reusable domain guidelines and general principles.}
    \label{fig:evolution-rounds}
    \vspace{-14pt}
\end{figure*}

We track validation quality and memory growth over five evolution rounds to assess the cumulative benefit of design expertise evolution.
The evolution curve shows that validation quality increases steadily as the system accumulates verified refinement traces, while the memory expands from local sample strategies into higher-level guidelines and principles.
This trend supports the role of the evolution mechanism as a reusable design-knowledge accumulator for later diagram generation.

\subsection{Qualitative and Human-Facing Analysis}

\begin{figure}[t]
    \centering
    \includegraphics[width=\linewidth]{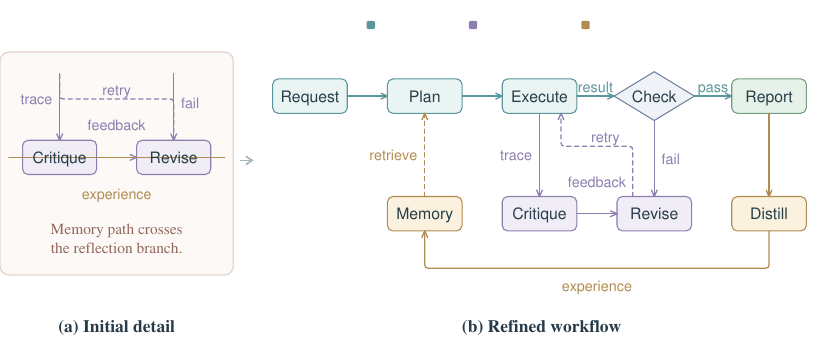}
    \caption{Localized refinement of a research-assistant workflow. Rerouting the memory return clears the reflection branch while preserving all nine nodes and eleven directed connections.}
    \label{fig:refinement-case-main}
    \vspace{-8pt}
\end{figure}

Figure~\ref{fig:refinement-case-main} shows a memory-return edge crossing Critique and Revise in a research-assistant workflow.
Rerouting this edge around the reflection branch clears the intersections while preserving the nodes, labels, and directed connections.
The complete workflow and cross-paradigm generation comparison appear in Appendix~\ref{app:case:refinement} and Appendix~\ref{app:case:diagram}.
The web application exposes the same manifest as object-level canvas elements.
Users can create or reopen projects, access design-memory controls, and adjust element-level properties through direct manipulation.
A selection-scoped design assistant accepts natural-language requests for local edits, allowing users to combine automated refinement with manual changes on the same canvas (Appendix~\ref{app:human-study}).

\subsection{Human Study}
\label{sec:human-study}

To complement automatic evaluation, we conducted a user study with 36 participants who had prior experience in diagram authoring.
Participants used the web application for their own diagramming tasks at least ten times within a day before completing eight questionnaire items on a five-point Likert scale (1 = strongly disagree, 5 = strongly agree).
The items cover personalized memory and editable refinement; the protocol is provided in Appendix~\ref{app:human-study}.

\begin{figure}[t]
    \centering
    \includegraphics[width=\linewidth]{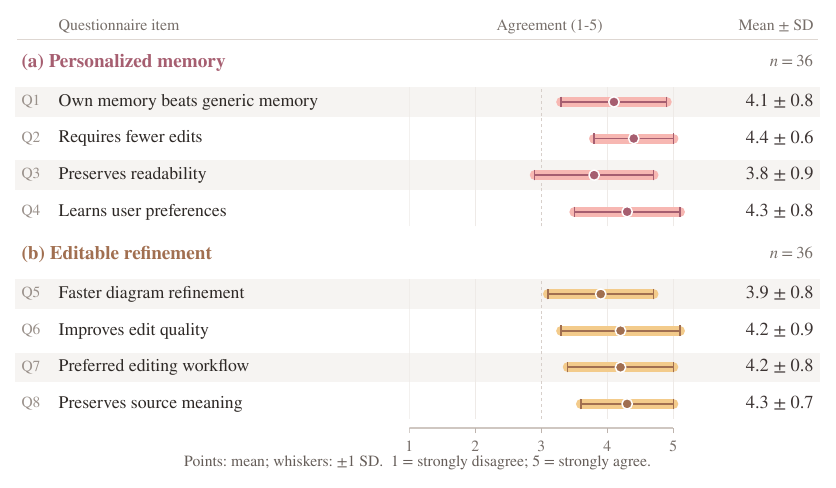}
    \caption{Self-reported user experience with EvoDesign ($n=36$). Points show mean agreement and horizontal whiskers show one standard deviation. Panels summarize (a) personalized memory and (b) editable refinement.}
    \label{fig:human-study-summary}
    \vspace{-8pt}
\end{figure}

As shown in Figure~\ref{fig:human-study-summary}, participants gave the highest rating to requiring fewer edits ($4.4 \pm 0.6$).
They also reported that the system learned their preferences ($4.3 \pm 0.8$), preserved source meaning during refinement ($4.3 \pm 0.7$), and provided a preferred editing workflow ($4.2 \pm 0.8$).
These ratings indicate positive perceived personalization and editing quality.

\Needspace{8\baselineskip}
\section{Conclusion}
In this paper, we introduced EvoDesign, an agentic framework designed to bridge the representation gap in automated diagramming. By generating diagrams through a canvas-based schema rather than static pixels or brittle code, EvoDesign enables a fluid transition between autonomous AI generation and human-centric UI intervention. The framework addresses multi-layer diagramming constraints through a coordinated multi-agent pipeline and a design knowledge evolution mechanism that can organize verified refinement traces into reusable expertise.
To support rigorous research in this paradigm, we define CanvasBench, a dataset construction process and evaluation protocol focused on object-level editability and cognitive utility. Experiments across pixel-based, code-based, agentic, one-shot schema, and canvas-native methods show that EvoDesign provides a balanced advantage on visual organization and cognitive usability while remaining competitive on content fidelity. In a user study with 36 participants, users reported positive experiences with personalized memory and editable refinement.

\clearpage

\bibliography{main}
\bibliographystyle{plainnat}

\appendix

\section{Further Discussion}
\label{app:further-discussion}

\subsection{Limitations}
\label{app:limitations}

\textbf{Canvas Schema Expressivity.}
The canvas schema supports general-purpose diagrams such as flowcharts, entity-relationship models, concept maps, and system architectures.
Support is more limited for dense scientific visualizations, complex mathematical notation, and artistic illustrations.

\textbf{Cold-Start Behavior and Cross-Model Generalization.}
Memory quality depends on the number and diversity of accumulated refinement traces.
Cold-start stability and transfer across backbone model families require further study.

\textbf{Benchmark Input Construction and Selection.}
CanvasBench's reference-derived inputs can specify structure more explicitly than independently authored requests, and its editability filter favors structured diagrams.
Less structured artifacts and naturally authored requests are therefore underrepresented.

\textbf{Computational Cost.}
Each diagram generation incurs multiple LLM calls for semantic structure, visual style, spatial layout, and refinement.
The knowledge evolution phase adds additional distillation calls per sample.
These calls increase interactive latency; smaller-model routing and output caching could reduce the cost.

\subsection{Broader Impacts}

EvoDesign lowers the barrier to producing structured visual artifacts, which can help non-experts communicate complex information across education, business, science, and engineering workflows.
Its canvas-based representation also supports human-AI co-creation by keeping generated artifacts object-level editable rather than static raster images.
Incomplete or incorrectly parsed source content can produce misleading diagrams, making human review important.
CanvasBench also uses web-derived visual materials during dataset construction, so release and redistribution should respect source attribution, licensing, and privacy constraints.

\subsection{LLM Usage}

We use LLMs and VLMs as core components of the proposed system, including diagram manifest generation, diagnostic critique, localized refinement, evaluation protocol design, and design expertise distillation.
We also used general-purpose language models during paper development for editing assistance, grammar checking, and concise wording suggestions.

\subsection{Code and Artifacts}

Code and prompts are available at \url{https://github.com/GeminiLight/EvoDesign}.

\section{System Implementation Details}
\label{app:system-implement}

\subsection{Algorithmic Overview}
\label{app:algorithm}

Algorithm~\ref{alg:generation} summarizes the end-to-end diagram generation pipeline, including multi-agent coordination and the closed-loop refinement mechanism. Algorithm~\ref{alg:evolution} details the progressive expertise evolution cycle that distills verified traces into a maintained hierarchical design memory.

\begin{algorithm}[t]
\caption{Diagram Generation with Closed-Loop Refinement}
\label{alg:generation}
\begin{algorithmic}[1]
\REQUIRE Content $c$, optional requirements $r$, knowledge memory $\mathcal{M}$, max refinement iterations $T_{\max}$ (set to 1 in the main experiments)
\ENSURE Rendered diagram artifact $\hat{\mathcal{D}}'$, refinement trace $\mathcal{T}$
\STATE Retrieve context-aware heuristics $\mathcal{K} \subset \mathcal{M}$ based on task metadata
\STATE $G \leftarrow A_{\text{str}}(\mathcal{K}_s, c, r)$ \hfill \COMMENT{Semantic Structure Agent}
\STATE $S \leftarrow A_{\text{sty}}(\mathcal{K}_{\text{sty}}, c, r, G)$ \hfill \COMMENT{Visual Style Agent}
\STATE $L \leftarrow A_{\text{lay}}(\mathcal{K}_{\text{lay}}, c, r, G, S)$ \hfill \COMMENT{Spatial Layout Agent}
\STATE $\mathcal{D} \leftarrow (G, S, L)$ \hfill \COMMENT{Compose canvas manifest}
\STATE $\hat{\mathcal{D}} \leftarrow \textsc{Render}(\mathcal{D})$ \hfill \COMMENT{Render to canvas artifact}
\STATE $\mathcal{T} \leftarrow \emptyset$ \hfill \COMMENT{Refinement evidence trace}
\FOR{$t = 1$ \textbf{to} $T_{\max}$}
    \STATE $\mathcal{F} \leftarrow \textsc{Diagnose}(\hat{\mathcal{D}}, c)$ \hfill \COMMENT{Objective checks + VLM critique}
    \IF{$\mathcal{F} = \emptyset$}
        \STATE \textbf{break} \hfill \COMMENT{No defects found}
    \ENDIF
    \STATE $\hat{\mathcal{D}}' \leftarrow A_{\text{ref}}(\hat{\mathcal{D}}, \mathcal{F})$ \hfill \COMMENT{Localized Refiner Agent}
    \STATE $\mathcal{F}' \leftarrow \textsc{Diagnose}(\hat{\mathcal{D}}', c)$
    \STATE $\Delta_{\text{score}}^{(t)} \leftarrow \textsc{Score}(\hat{\mathcal{D}}') - \textsc{Score}(\hat{\mathcal{D}})$
    \STATE Record evidence $e^{(t)}=(\mathcal{F}, a^{(t)}, \mathcal{F}', \Delta_{\text{score}}^{(t)})$
    \STATE $\mathcal{T} \leftarrow \mathcal{T} \cup \{e^{(t)}\}$
    \STATE $\hat{\mathcal{D}} \leftarrow \hat{\mathcal{D}}'$
\ENDFOR
\RETURN $\hat{\mathcal{D}}, \mathcal{T}$
\end{algorithmic}
\end{algorithm}

\begin{algorithm}[t]
\caption{Progressive Design Expertise Evolution}
\label{alg:evolution}
\begin{algorithmic}[1]
\REQUIRE Training samples $\{I_i\}_{i=1}^{N}$, evolution rounds $R$, strategy threshold $\tau=0.1$, knowledge memory $\mathcal{M} = \{\mathcal{K}^s, \mathcal{K}^g, \mathcal{K}^p\}$
\ENSURE Updated memory $\mathcal{M}$
\FOR{round $r = 1$ \textbf{to} $R$}
    \STATE Sample batch $\mathcal{B} \subset \{I_i\}_{i=1}^{N}$
    \FOR{each task instance $I \in \mathcal{B}$}
        \STATE Retrieve $\mathcal{K} \subset \mathcal{M}$ based on task metadata of $I$
        \STATE $\mathcal{T}_{\mathrm{vert}} \leftarrow$ Run Algorithm~\ref{alg:generation} on $I$ and collect vertical refinement evidence
        \STATE $\{I'_j\}_{j=1}^{m} \leftarrow$ Generate requirement or context variants for $I$
        \STATE $\mathcal{T}_{\mathrm{horiz}} \leftarrow \emptyset$
        \FOR{each variant $I'_j$}
            \STATE $\mathcal{T}_{\mathrm{horiz}} \leftarrow \mathcal{T}_{\mathrm{horiz}} \cup$ Run Algorithm~\ref{alg:generation} on $I'_j$
        \ENDFOR
        \STATE $\mathcal{T}_I \leftarrow \mathcal{T}_{\mathrm{vert}} \cup \mathcal{T}_{\mathrm{horiz}}$
        \STATE $\mathcal{C}_I \leftarrow$ Extract candidate strategies from $\mathcal{T}_I$ with evidence links
        \FOR{each candidate strategy $c$ with supporting evidence $E_c \subset \mathcal{T}_I$}
            \STATE $q_c \leftarrow |E_c|^{-1}\sum_{e\in E_c}\Delta_{\mathrm{score}}(e)$ \hfill \COMMENT{Mean linked-step improvement}
            \IF{$q_c > \tau$}
                \STATE $k^s \leftarrow$ Distill a sample strategy with precondition, action, effect, tags, confidence $q_c$, and provenance $E_c$
                \STATE $\mathcal{K}^s \leftarrow \mathcal{K}^s \cup \{k^s\}$
            \ELSIF{$q_c < -\tau$}
                \STATE Record the trace as low-utility evidence for pruning or demotion
            \ENDIF
        \ENDFOR
    \ENDFOR
    \STATE Group related strategies by agent role, diagram type, domain, and action template
    \STATE Promote strategy groups supported by $\geq 5$ traces across $\geq 3$ domains into $\mathcal{K}^g$
    \STATE Promote guideline groups retaining this support and spanning $\geq 2$ diagram types into $\mathcal{K}^p$
    \STATE $\mathcal{M} \leftarrow \textsc{Score}(\mathcal{M})$
    \STATE $\mathcal{M} \leftarrow \textsc{Prune}(\mathcal{M})$
    \STATE $\mathcal{M} \leftarrow \textsc{Merge}(\mathcal{M})$
    \STATE $\mathcal{M} \leftarrow \textsc{Promote}(\mathcal{M})$
    \STATE $\mathcal{M} \leftarrow \textsc{Demote}(\mathcal{M})$
\ENDFOR
\RETURN $\mathcal{M}$
\end{algorithmic}
\end{algorithm}

\subsection{Specification Dimension Summarization of Agents}
\label{app:implementation:spec_dimensions}
The Semantic Structure Agent specifies the diagram type, visual rationale, primary flow, and grouping before instantiating the semantic graph.

\begin{itemize}
\item \textit{Selected Diagram Type} represents the precise visualization archetype chosen from the knowledge base, such as a Flowchart, Mindmap, and Block Diagram, to align the structural skeleton with the underlying patterns of the text.
\item \textit{Visual Rationale} establishes a concise mental blueprint that governs the narrative scope and serves as a semantic filter to exclude high-entropy noise.
\item \textit{Primary Flow Strategy} dictates the global directional logic such as linear propagation to ensure the synthesized topology remains legible and logically sound.
\item \textit{Grouping Logic} defines the boundaries of organizational containers to aggregate related entities and maintain relational proximity within the semantic structure.
\end{itemize}

The Visual Style Agent specifies visual tone, hierarchy, color, shapes, and connector styles.

\begin{itemize}
    \item \textit{Visual Tone Principle} determines the overall artistic character of the diagram by utilizing specific font and line styles to create a professional atmosphere.
    \textit{Hierarchy Principle} formulates a strategy for visual prioritization by managing the sequential rank index of elements to ensure a clear and organized depth of information.
    \item \textit{Color Principle} maps distinct palettes to elements based on their categorical meaning to enhance thematic coherence across the entire canvas.
    \item \textit{Shape Principle} selects geometric primitives that visually communicate the functional role of each element to ensure intuitive recognition.
    \item \textit{Connection Principle} prescribes the visual defaults for relational edges including arrowhead styles and stroke colors to minimize visual clutter.
\end{itemize}

The Spatial Layout Agent specifies flow direction, spacing, alignment, layering, and nested organization.

\begin{itemize} 
\item \textit{Flow Strategy Principle} governs the global arrangement of components based on relational dependencies to determine the optimal directional orientation.

\item \textit{Spatial Breathing Principle} regulates the margin and padding logic by employing a progressive tightening strategy to ensure sufficient whitespace around complex groups.

\item \textit{Alignment Balance Principle} organizes parallel or sibling items into balanced grids to account for the physical footprint of text-heavy shapes.
\item \textit{Depth Layering Principle} manages the vertical stacking order to ensure that nested children are rendered on top of parent backgrounds without occlusion.

\item \textit{Structural Rhythm Principle} optimizes space utilization by alternating between horizontal flows for outer modules and vertical lists for inner details.
\end{itemize}

\subsection{Predefined Diagram Types in EvoDesign}
We manually curated twenty-one diagram types specifically tailored for canvas editability. Table~\ref{tab:diagram_types_appendix} summarizes the key characteristics and optimal applications for these selected types.
\label{app:implementation:diagram_typs}
\begin{table}[htbp]
    \centering
    \caption{Comprehensive Summary of Diagram Types.}
    \label{tab:diagram_types_appendix}
    \small 
    
    \begin{tabularx}{\linewidth}{p{3.5cm} >{\RaggedRight\arraybackslash}X >{\RaggedRight\arraybackslash}X}
        \toprule
        \textbf{Chart Type} & \textbf{Best For} & \textbf{Key Characteristics} \\
        \midrule
        \textit{Architecture Diagram} & System design, software structure & Modules, components, and interfaces \\
        \textit{Class Diagram} & OOP modeling, database schemas & Classes, attributes, methods, inheritance \\
        \textit{Concept Map} & Knowledge organization, learning & Nodes connected by labeled relationships \\
        \textit{Data Flow Diagram} & System information flow analysis & Processes, data stores, external entities \\
        \textit{Entity Relationship} & Database design, modeling data & Entities, attributes, relationships (1:N, M:N) \\
        \textit{Fishbone Diagram} & Root cause analysis & Spine (problem) with ribs (causes) \\
        \textit{Flowchart} & Process steps, decision logic & Sequential shapes with directional arrows \\
        \textit{Funnel Chart} & Sales pipelines, conversion rates & Tapering stages showing reduction \\
        \textit{Gantt Chart} & Project scheduling, timelines & Horizontal bars representing tasks over time \\
        \textit{Hybrid Diagram} & Complex multi-view systems & Combination of elements from various diagrams \\
        \textit{Matrix Chart} & Multi-variable comparisons & Grid structure intersecting data points \\
        \textit{Mind Map} & Brainstorming, ideation & Central topic with radiating sub-branches \\
        \textit{Org Chart} & Hierarchy, reporting lines & Tree structure of roles and ranks \\
        \textit{Pyramid Chart} & Hierarchical levels, priority & Stacked triangular layers \\
        \textit{Sequence Diagram} & Interaction ordering, messaging & Vertical lifelines with time-based messages \\
        \textit{State Diagram} & System states, transitions & Nodes triggered by events \\
        \textit{Swimlane Diagram} & Cross-functional workflows & Parallel lanes distinguishing responsibilities \\
        \textit{SWOT Analysis} & Strategic planning & 2x2 matrix (Strengths, Weaknesses, etc.) \\
        \textit{Timeline} & Chronological history, events & Linear axis marked with dates/milestones \\
        \textit{Tree Diagram} & Decomposition, probability & Root node branching into hierarchical children \\
        \textit{Venn Diagram} & Set relationships, logic overlaps & Overlapping circles sharing attributes \\
        \bottomrule
    \end{tabularx}
\end{table}
\subsection{Canvas Schema Space in Excalidraw Library}
\label{app:implementation:excalidraw}
The Excalidraw schema represents diagrams through styling and layout properties. Table~\ref{tab:excalidraw_action_space} lists shapes, colors, strokes, and typography; Table~\ref{tab:layout_action_space} lists coordinates, dimensions, and layer order. Agents instantiate these properties to produce editable canvas objects.

\subsection{Auxiliary Layout Support in EvoDesign}
\label{app:implementation:layout_support}

\subsubsection{Geometric Size Estimator for Layout Agent}
\label{app:implementation:estimation}
The Geometric Size Estimator estimates minimum node bounds from text wrapping, font properties, padding, and icon space. Dimensions are rounded up to integer values before recursive placement.
\begin{table*}[htbp]
\small
\centering
\caption{Styling-related properties and categorical parameter spaces of \texttt{excalidraw} shape elements}
\label{tab:excalidraw_action_space}
\begin{tabular}{@{}llp{10cm}@{}}
\toprule
\textbf{Category} & \textbf{Property} & \textbf{Candidate Parameters Values} \\ 
\midrule
\textbf{Color} & \texttt{TLColor} & black, grey, light-violet, violet, blue, light-blue, yellow, orange, green, light-green, light-red, red, white \\ \addlinespace
\textbf{Geometry} & \texttt{TLGeo} & rectangle, ellipse, triangle, diamond, pentagon, hexagon, octagon, star, cloud, arrowLeft, arrowRight, arrowUp, arrowDown \\ \addlinespace
\textbf{Size} & \texttt{TLSize} & s, m, l, xl \\ \addlinespace
\textbf{Fill Style} & \texttt{TLFill} & none, semi, solid, pattern \\ \addlinespace
\textbf{Dash Style} & \texttt{TLDash} & draw, solid, dashed, dotted \\ \addlinespace
\textbf{Typography} & \texttt{TLFont} & draw, sans, serif, mono \\ \addlinespace
\bottomrule
\end{tabular}
\end{table*}

\begin{table*}[htbp]
\small
\centering
\caption{Layout-related properties and discretized parameter spaces of  \texttt{excalidraw} shape elements}
\label{tab:layout_action_space}
\begin{tabular}{@{}llp{6cm}@{}}
\toprule
\textbf{Category} & \textbf{Property} & \textbf{Restricted Parameter Space} \\ \midrule
\textbf{Position} & \texttt{x, y} & Positive Integers \\ \addlinespace
\textbf{Dimensions} & \texttt{w, h} & Positive Integers \\ \addlinespace
\textbf{Rotation} & \texttt{rotation} & Integer Degrees \\ \addlinespace
\textbf{Alignment} & \texttt{align} & \texttt{start}, \texttt{middle}, \texttt{end} \\ \addlinespace
\textbf{Layering} & \texttt{index} & Sequential Rank Index \\ 
\bottomrule
\end{tabular}
\end{table*}

\section{Details of Evaluation Settings}
\label{app:experiment}

\textbf{Implementation Details}.
We utilize the popular \texttt{excalidraw}\footnote{\url{https://excalidraw.com/}} library as the underlying canvas engine.
See Appendix~\ref{app:implementation:excalidraw} for its schema description.
The agentic framework is orchestrated using \textsc{LangGraph} and powered by VLM backbones.
For the design expertise memory, we use a vector store and embedding model to retrieve relevant expertise items by semantic similarity.
Unless otherwise stated, generation uses a temperature of 0.7, one localized refinement pass, and top-$k$ retrieval for each knowledge type.

\textbf{Data Splits and Evolution Settings}.
CanvasBench defines a balanced held-out test set and a remaining pool for knowledge evolution.
The main tables report a balanced evaluation subset spanning the diagram types in CanvasBench.
For memory evolution, each round selects samples for iterative exploration with requirement variants, refinement traces, promotion thresholds, and maintenance operations.
Detailed prompts are provided in our codebase\footnote{\url{https://github.com/GeminiLight/EvoDesign}}.

\textbf{Baseline Settings}.
The baselines cover four paradigms:
\textit{Pixel-based Generation}, where image-generation models directly produce raster diagrams;
\textit{Code-based Synthesis}, where LLMs generate \LaTeX{} TikZ code~\cite{diagram-cvpr-2025-text2diagram};
\textit{One-shot Schema Generation}, where LLMs directly generate the canvas schema manifest; and
\textit{Agentic Diagram Systems}, where existing multi-agent diagramming workflows are adapted to the same task format.
For fair comparison, Table~\ref{tab:main_results} separates method names from backbone models and evaluates non-pixel baselines under GPT-5.5, Gemini-3.1-Pro, DeepSeek-V4-Pro, and DeepSeek-V4-Flash whenever the method can use a backbone VLM.

\subsection{Multi-Seed and Cross-Judge Checks}
\label{app:robustness}

\textbf{Generation variability.}
We repeated the main comparison with three independent generation seeds, using one representative model per paradigm and GPT-5.5 for the LLM-based methods.
Table~\ref{tab:seed-robustness} reports the mean and standard deviation across seeds for each metric on the five-point scale.
The standard deviations describe variability due to generation seeds.
EvoDesign has higher mean scores than TikZ and one-shot prompting on all ten metrics; pixel generation has higher means on Visual Aesthetics, Space Balance, and Cognitive Ease.

\begin{table}[H]
    \centering
    \caption{Three-seed comparison: mean $\pm$ standard deviation across independent generation seeds. Pixel generation uses GPT-4o-Image; TikZ, prompting, and EvoDesign use GPT-5.5. The two blocks show all ten CanvasBench metrics.}
    \label{tab:seed-robustness}
    \small
    \setlength{\tabcolsep}{4pt}
    \renewcommand{\arraystretch}{1.12}
    \begin{tabular}{lccccc}
    \toprule
    Method & $C_{CF}$ & $C_{CL}$ & $C_{SR}$ & $V_{VA}$ & $V_{SC}$ \\
    \midrule
    Pixel & $2.14 \pm 0.17$ & $2.31 \pm 0.13$ & $2.47 \pm 0.11$ & $3.77 \pm 0.10$ & $3.66 \pm 0.08$ \\
    TikZ & $3.37 \pm 0.09$ & $3.45 \pm 0.12$ & $3.55 \pm 0.07$ & $2.71 \pm 0.16$ & $2.54 \pm 0.13$ \\
    Prompting & $3.91 \pm 0.11$ & $3.53 \pm 0.09$ & $3.58 \pm 0.12$ & $2.50 \pm 0.15$ & $2.86 \pm 0.12$ \\
    \textbf{EvoDesign} & $4.09 \pm 0.08$ & $4.07 \pm 0.10$ & $4.15 \pm 0.04$ & $3.58 \pm 0.13$ & $3.81 \pm 0.05$ \\
    \midrule
    Method & $V_{FC}$ & $V_{SB}$ & $G_{CE}$ & $G_{CA}$ & $G_{SE}$ \\
    \midrule
    Pixel & $3.62 \pm 0.14$ & $3.60 \pm 0.09$ & $4.01 \pm 0.10$ & $2.32 \pm 0.16$ & $2.88 \pm 0.12$ \\
    TikZ & $3.29 \pm 0.15$ & $2.89 \pm 0.14$ & $3.81 \pm 0.12$ & $3.01 \pm 0.10$ & $2.82 \pm 0.15$ \\
    Prompting & $2.55 \pm 0.18$ & $2.03 \pm 0.16$ & $2.77 \pm 0.13$ & $3.00 \pm 0.10$ & $2.45 \pm 0.14$ \\
    \textbf{EvoDesign} & $3.68 \pm 0.11$ & $3.56 \pm 0.09$ & $3.94 \pm 0.10$ & $3.48 \pm 0.07$ & $3.11 \pm 0.12$ \\
    \bottomrule
    \end{tabular}
\end{table}

\textbf{Judge agreement.}
For a separate 100-item check, we compared Kimi-K3 and GPT-5.2 judgments using per-item Spearman rank correlations for each metric.
Correlations range from $0.58$ for Cognitive Ease to $0.74$ for Content Fidelity, with a mean difference of $0.05$ in the aggregate five-point score.
The metric-wise correlations are $0.74$, $0.63$, $0.69$, $0.70$, $0.66$, $0.61$, $0.72$, $0.58$, $0.71$, and $0.64$ for $C_{CF}$, $C_{CL}$, $C_{SR}$, $V_{VA}$, $V_{SC}$, $V_{FC}$, $V_{SB}$, $G_{CE}$, $G_{CA}$, and $G_{SE}$, respectively.

\section{Details of CanvasBench Dataset and Benchmark}
\label{app:benchmark}

\subsection{Description of Data Collection Pipeline}
\label{app:benchmark:collection}
As illustrated in Figure~\ref{fig:benchmark}, we implement a three-stage pipeline to filter and synthesize diagram-instruction pairs using diagram editability as the selection criterion.

\textit{(1) Structured Image Search.}
To ensure broad coverage of real-world scenarios, we construct a search matrix intersecting two axes. We manually curated 21 diagram types (e.g., Flowcharts, Mind Map) specifically tailored for canvas editability, while adopting the 30 vertical domains (e.g., Finance, Healthcare) from the established MMMU benchmark~\cite{benchmark-cvpr-2024-mmmu-benchamrk,benchmark-arxiv-2025-video-mmmu} to guarantee content diversity. The full taxonomy is detailed in Appendix~\ref{app:benchmark:taxonomy}. 
We utilize the Google Images API to retrieve candidate queries based on these enumerated tuples, i.e., ``\texttt{diagram\_type} diagram of \texttt{vertical\_domain}''.

\textit{(2) Diagram Editability Validation.}
We use Qwen3-VL-Max~\cite{llm-arxiv-2025-qwen3vl} to filter out hand-drawn sketches, complex data visualizations, photo infographics, and blurred documents, retaining diagrams whose structure can be represented as editable objects.
Section~\ref{sec:benchmark:collection} reports the independent annotation and deduplication protocol.

\textit{(3) Content \& Intent Synthesis.}
Since raw web data lacks explicit user instructions and structured captions, we employ a data completion strategy to synthesize the missing information. We instruct Qwen3-VL-Max to jointly analyze both the diagram image and its associated webpage context to reverse-engineer generation. This process generates both a dense diagram description grounding the visual topology and a canonical user query (e.g., ``Design a deployment pipeline...'') representing the likely user intent.
Consequently, this step transforms the raw data into (image, query, content) triplets.

\subsection{Taxonomy for Diagram Image Retrieval}
\label{app:benchmark:taxonomy}
Table~\ref{tab:taxonomy_retrieval} lists the 21 diagram types and 30 vertical domains combined to construct the retrieval queries.

\definecolor{boxheader}{RGB}{235, 240, 255}
\definecolor{boxborder}{RGB}{70, 130, 180}

\begin{table}[htbp] 
    \centering
    \caption{Detailed Taxonomy for Diagram Image Retrieval. The taxonomy intersects 21 structural types with 30 semantic domains.} 
    \label{tab:taxonomy_retrieval}
\begin{tcolorbox}[
    enhanced,
    title={\textbf{Dual-axis Taxonomy for Diagram Image Retrieval}},
    colframe=boxborder,
    colback=white,
    colbacktitle=boxheader,
    coltitle=black,
    attach boxed title to top center={yshift=-2mm},
    boxrule=0.5mm,
    arc=2mm,
    drop fuzzy shadow,
    top=1.2em, bottom=1em, left=1em, right=1em
]
\textbf{\textcolor{boxborder}{Axis 1: Structural Diversity (21 Diagram Types)}} \\
\textit{\small Selection criterion: Diagram editability.}
\vspace{0.5em}

\begin{multicols}{3}
\small
\begin{itemize}[leftmargin=*, nosep]
    \item Architecture Diagram
    \item Class Diagram
    \item Concept Map
    \item Data Flow Diagram
    \item Entity Relationship (ER)
    \item Fishbone Diagram
    \item Flowchart
    \item Funnel Chart
    \item Gantt Chart
    \item Hybrid Diagram
    \item Matrix Chart
    \item Mind Map
    \item Org Chart
    \item Pyramid Chart
    \item Sequence Diagram
    \item State Diagram
    \item Swimlane Diagram
    \item SWOT Analysis
    \item Timeline
    \item Tree Diagram
    \item Venn Diagram
\end{itemize}
\end{multicols}

\vspace{0.1cm}

\textbf{\textcolor{boxborder}{Axis 2: Semantic Diversity (30 Vertical Domains)}} \\
\textit{\small Taxonomy Source: Adopted from MMMU Benchmark~\cite{benchmark-cvpr-2024-mmmu-benchamrk}.}
\vspace{0.5em}

\begin{itemize}[style=multiline, leftmargin=3.8cm, parsep=0.3em]
\small
    \item[Art \& Design] Art, Design, Music, Art Theory
    \item[Business] Accounting, Economics, Finance, Management, Marketing
    \item[Science] Biology, Chemistry, Geography, Math, Physics
    \item[Health \& Medicine] Basic Medicine Science, Clinical Medicine, Diagnostics, Pharmacy, Public Health
    \item[Humanities] History, Literature, Psychology, Sociology
    \item[Tech \& Eng.] Agriculture, Architecture, CS, Electronics, Energy, Materials, Mechanical Engineering
\end{itemize}
\label{tab:taxonomy-retrieval}
\end{tcolorbox}
\end{table}

\subsection{Further Dataset Analysis}
\label{app:benchmark:analysis}

We analyze description length and domain coverage to characterize CanvasBench.

\begin{figure*}[t]
    \centering
    \includegraphics[width=0.98\linewidth]{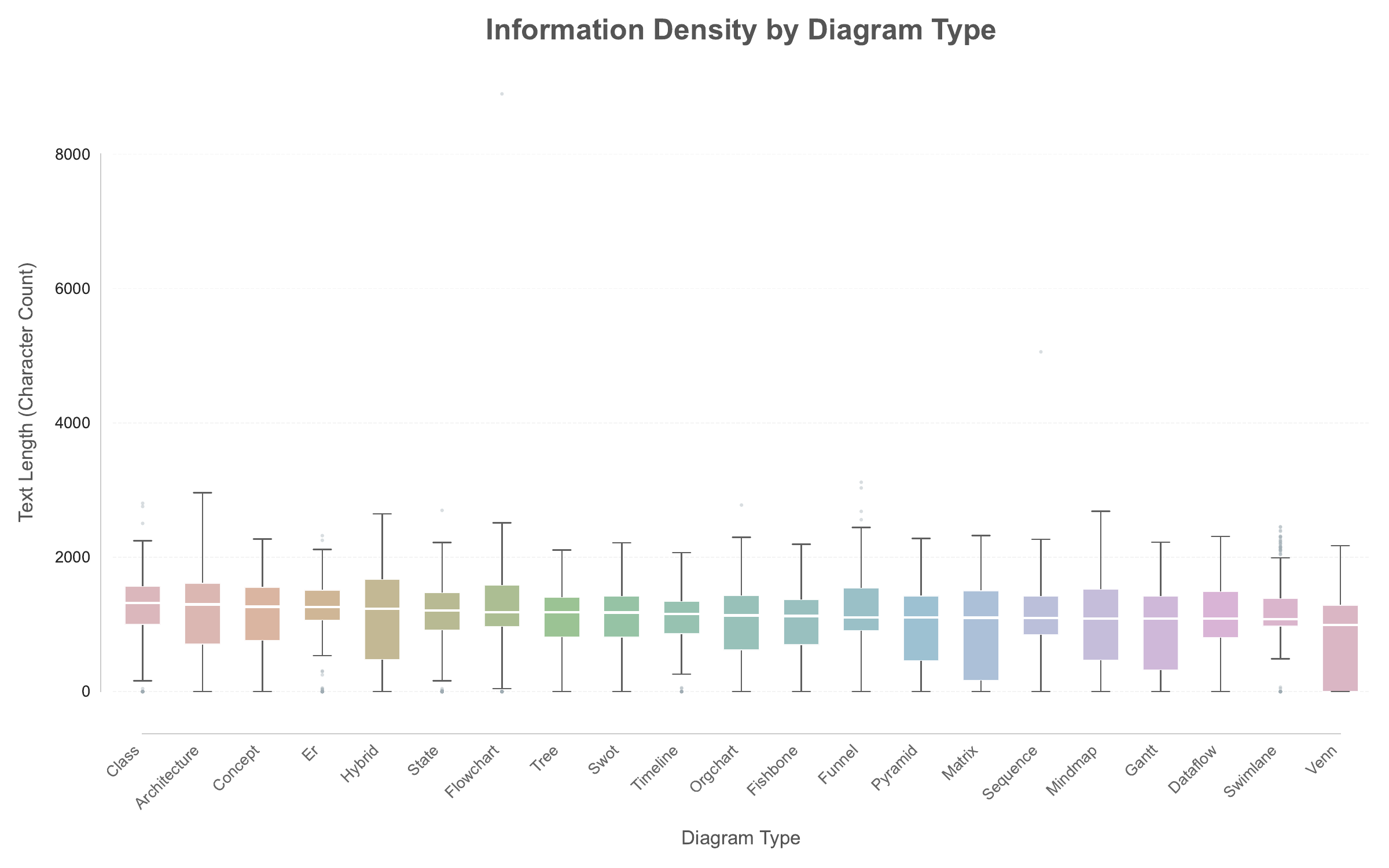}
    \caption{Distribution of description character counts by diagram type.}
    \label{fig:density}
\end{figure*}

\paragraph{Semantic Complexity and Information Density.}
Constructing high-fidelity diagrams requires processing not only topological structures but also dense textual information. We quantify this complexity by analyzing the character count distribution across the 21 diagram types, as illustrated in Figure~\ref{fig:density}. The results demonstrate that CanvasBench maintains a high level of information density, with the median text length for most categories exceeding 1,000 characters. Notably, knowledge-intensive types such as \textit{Concept Maps}, \textit{Mind Maps}, and \textit{Entity Relationship (ER)} diagrams exhibit substantial variance and higher upper quartiles, reflecting their role in organizing complex, multi-faceted concepts. Even structurally rigid types like \textit{Sequence Diagrams} maintain significant textual volume, underscoring the necessity for models to possess strong long-context understanding capabilities alongside layout generation skills.

\begin{figure*}[t]
    \centering
    \includegraphics[width=0.58\linewidth]{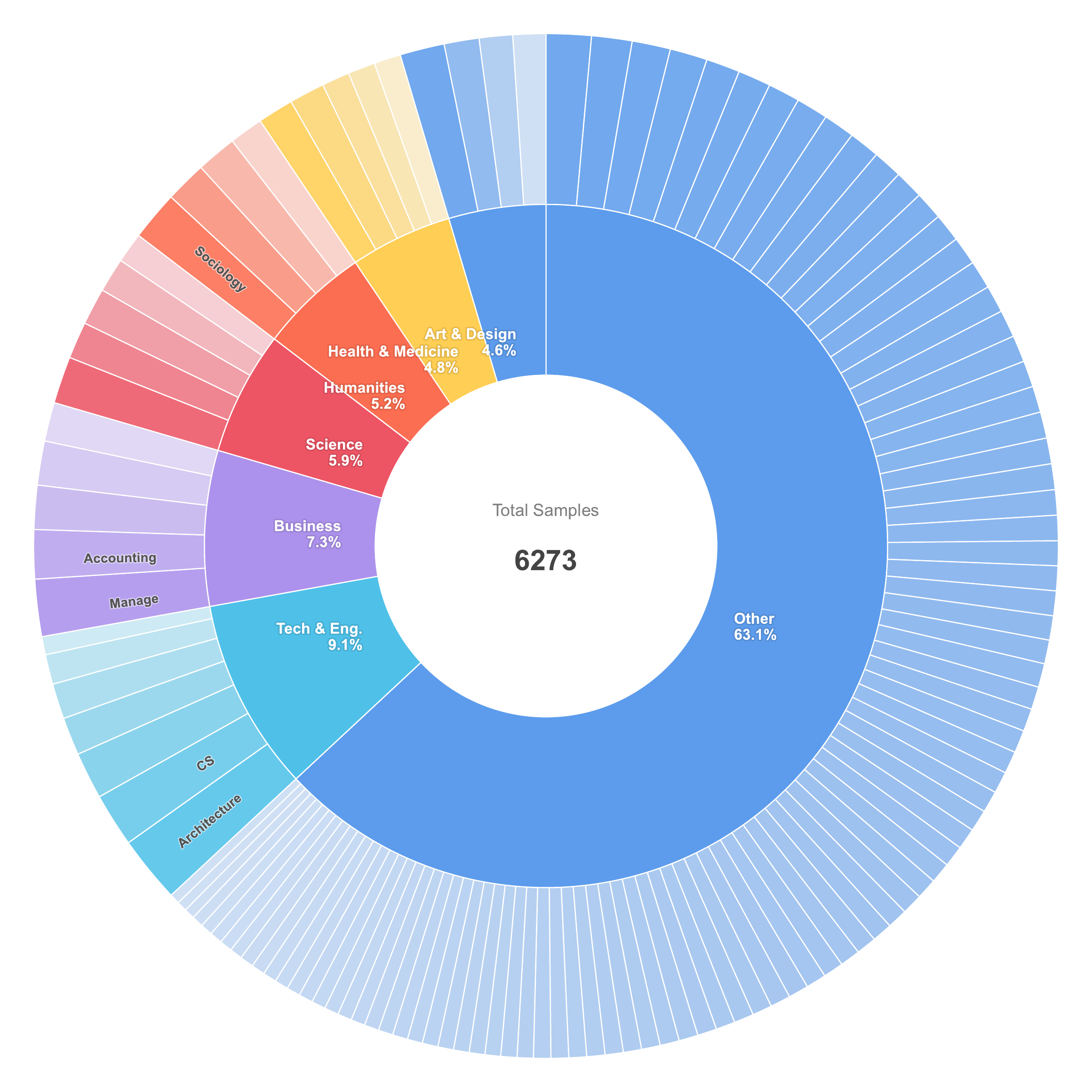}
    \caption{Vertical domain coverage in the current corpus. The chart summarizes the 30 MMMU-derived vertical domains used for CanvasBench construction.}
    \label{fig:domain}
\end{figure*}

\paragraph{Domain Coverage.}
Figure~\ref{fig:domain} presents the vertical-domain coverage of the current CanvasBench corpus. The dataset follows the 30 MMMU-derived vertical domains used during retrieval, which provide broad coverage of terminology and logical conventions for knowledge evolution, while the held-out test set remains balanced by diagram type.

\subsection{Examples of Diagram Data}
\label{app:benchmark:examples}

\begin{figure*}[t]
    \centering
    \includegraphics[width=0.98\linewidth]{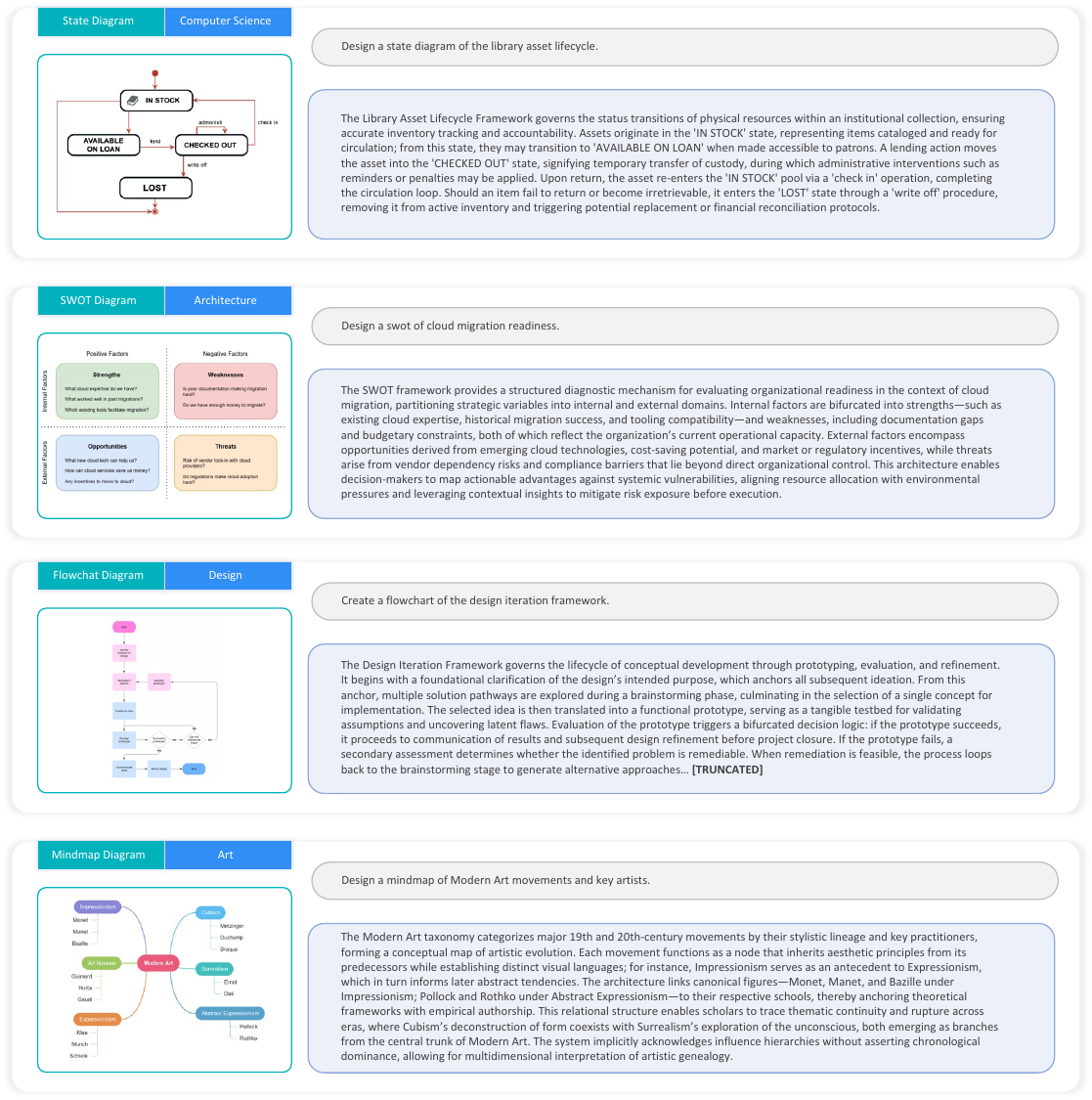}
    \caption{Data examples in CanvasBench.}
    \label{fig:examples}
\end{figure*}

To provide a concrete visualization of the diversity and quality of CanvasBench, we present several representative samples in Figure~\ref{fig:examples}. These examples showcase the benchmark's coverage across various structural types, including state diagrams, SWOT analysis, flowcharts, and mind maps, and their corresponding vertical domains such as Computer Science, Architecture, Design, and Art.
Each data point in the benchmark is a curated triplet consisting of a high-fidelity diagram image, a canonical user query, and a dense textual description. We can observe that

\begin{itemize}
    \item \textit{Structural Fidelity}: The examples demonstrate complex topological features, such as the cyclical transitions in the library asset lifecycle (State Diagram) and the hierarchical branching of artistic movements (Mind Map).
    \item \textit{Semantic Richness}: The associated source texts (right-hand panels) exhibit high information density, typically exceeding 1,000 characters. This requires models to not only extract entities but also to synthesize latent relationships into a coherent visual layout.
    \item \textit{Diagram Editability}: The samples contain object-based structures suitable for editable nodes, edges, and groups.
\end{itemize}

\subsection{Definitions and Prompts of Evaluation Metrics}
\label{app:benchmark:metrics}

In this section, we provide the detailed definitions and scoring rubrics for the evaluation framework. The evaluation utilizes VLM-based judges to assign scores on a 5-point Likert scale across three primary axes.

\textbf{(a) Content Integrity Dimension ($\mathcal{C}$)}
Focuses on the semantic accuracy of the translation from source text to schema.
\begin{itemize}
    \item \textit{Content Fidelity ($C_{CF}$)}: Verifies if every visual element is factually supported by the source text and penalizes "hallucinated" entities or relationships.
    \item \textit{Concept Logic ($C_{CL}$)}: Evaluates structural correctness, ensuring parent-child hierarchies and edge directions follow logical dependency rules.
    \item \textit{Semantic Relevance ($C_{SR}$)}: Assesses information density, ensuring key concepts are covered (Recall) and node text remains concise rather than verbose.
\end{itemize}

\textbf{(b) Visual Presentation Dimension ($\mathcal{V}$)}
Assesses the rendering quality and adherence to professional design principles.
\begin{itemize}
    \item \textit{Visual Aesthetics ($V_{VA}$)}: Rates the overall professional appearance, composition harmony, and effective use of negative space.
    \item \textit{Styling Consistency ($V_{SC}$)}: Checks for a uniform "Design Language," ensuring font hierarchy, color themes, and stroke styles are consistent across similar element types.
    \item \textit{Flow Coherence ($V_{FC}$)}: Assesses the clarity of the reading path and penalizes ambiguous or "spaghetti" edge routing.
    \item \textit{Space Balance ($V_{SB}$)}: Evaluates if layout density is balanced across the canvas and identifies spatial conflicts such as overlaps.
\end{itemize}

\textbf{(c) Cognitive Utility Dimension ($\mathcal{G}$)}
Section~\ref{sec:benchmark:evaluation} defines the three cognitive metrics: Cognitive Ease ($G_{CE}$), Comprehension Accuracy ($G_{CA}$), and Self-contained Explanation ($G_{SE}$).
Table~\ref{tab:metrics_overview} lists their focus and scoring sub-items.

\begin{table}[htbp]
\caption{Overview of Evaluation Metrics}
\label{tab:metrics_overview}
\centering
\renewcommand{\arraystretch}{1.2}
\resizebox{\textwidth}{!}{
\begin{tabular}{c c c p{5.2cm} p{8.8cm}} 
\toprule
\multicolumn{1}{c}{\textbf{Dimension}} & \multicolumn{1}{c}{\textbf{Metric Name}} & \multicolumn{1}{c}{\textbf{Symbol}} & \multicolumn{1}{c}{\textbf{Focus}} & \multicolumn{1}{c}{\textbf{Sub-items}} \\
\midrule
\multirow{3}{*}{\shortstack{\textbf{Content} \\ \textbf{Integrity}}}
 & Content Fidelity & $\text{C}_{\textit{CF}}$ & Trustworthiness \& Fact-checking & Factual Consistency, Hallucination Rate, Evidence Traceability \\
 & Concept Logic & $\text{C}_{\textit{CL}}$ & Logical Structure \& Relationships & Edge Correctness, Relation Type Accuracy, Self-Consistency, Hierarchy Rationality \\
 & Semantic Relevance & $\text{C}_{\textit{SR}}$ & Information Density \& Alignment & Coverage, Conciseness, Context Alignment \\
\midrule
\multirow{4}{*}{\shortstack{\textbf{Visual} \\ \textbf{Presentation}}}
 & Visual Aesthetics & $\text{V}_{\textit{VA}}$ & Overall Look \& Feel & Composition \& Whitespace, Harmony, Professionalism \\
 & Styling Consistency & $\text{V}_{\textit{SC}}$ & Standardization \& Uniformity & Font Consistency, Line \& Shape Consistency, Color Theme Consistency \\
 & Flow Coherence & $\text{V}_{\textit{FC}}$ & Reading Path \& Navigation & Main Reading Direction, Low Path Ambiguity, Key Path Trackability \\
 & Space Balance & $\text{V}_{\textit{SB}}$ & Layout \& Density & Spatial Equilibrium, Density, Overlap \& Crossing \\
\midrule
\multirow{3}{*}{\shortstack{\textbf{Cognitive} \\ \textbf{Utility}}}
 & Cognitive Ease & $\text{G}_{\textit{CE}}$ & Mental Effort \& Friction & Subjective Ease, Reading Friction, Mental Load \\
 & Comprehension Accuracy & $\text{G}_{\textit{CA}}$ & Understanding \& QA Correctness & Factual QA, Structural QA, Interpretation Consistency \\
 & Self-contained Explanation & $\text{G}_{\textit{SE}}$ & Independence from Source & Self-containment, Narrative Completeness, Misinterpretation Risk \\
\bottomrule
\end{tabular}
}
\end{table}

\section{Qualitative Artifacts and Knowledge Examples}
\label{app:result analysis}

\subsection{Generated Diagram Artifacts}
\label{app:case:diagram}

\begin{figure*}[t]
    \centering
    \includegraphics[width=0.96\linewidth]{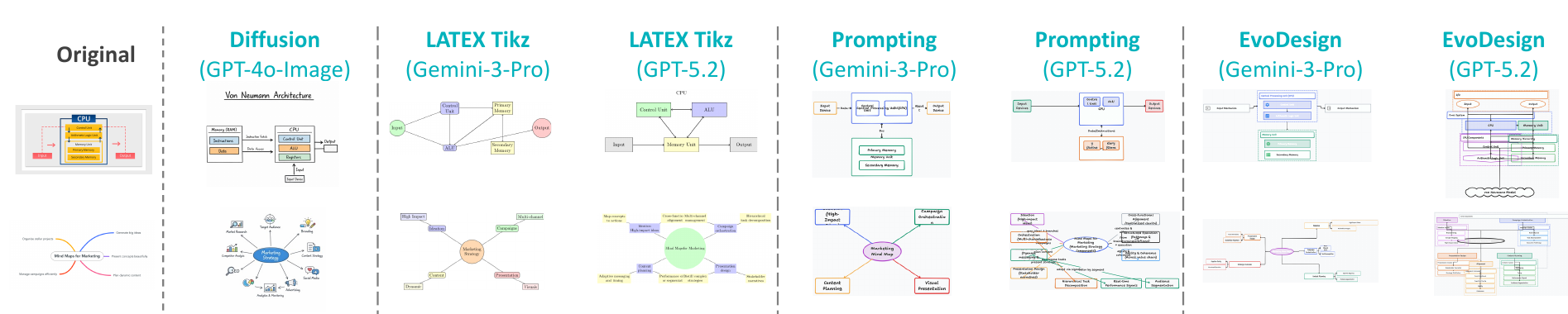}
    \caption{Representative diagram outputs from different generation paradigms on the same task.}
    \label{fig:mainpicture}
    \vspace{-12pt}
\end{figure*}

\Needspace{10\baselineskip}
\subsection{Geometric Refinement Example}
\label{app:case:refinement}
\begin{figure}[H]
    \centering
    \includegraphics[width=0.90\linewidth]{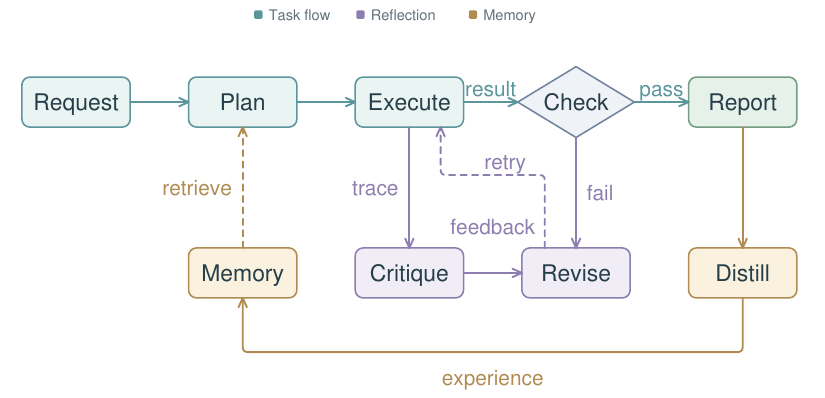}
    \caption{Refined research-assistant workflow. The task path, reflection loop, and memory return occupy separate routes while retaining all node positions and semantic connections.}
    \label{fig:iterations}
    \vspace{-12pt}
\end{figure}

Verification identifies two edge-node intersections: the Distill-to-Memory edge traverses Critique and Revise.
The refiner routes the memory return along the lower perimeter and separates the retry connector from the outgoing trace.
All nine nodes stay in place, and their labels and eleven directed connections are preserved.
Figure~\ref{fig:iterations} shows the complete refined workflow; Figure~\ref{fig:refinement-case-main} pairs the initial defect detail with the full result.

\subsection{Examples of Knowledge Items}
\label{app:case:knowledge}

Tables~\ref{tab:sample_strategy},~\ref{tab:domain guideline},~\ref{tab:general_principle} show example entries in the three-tier design memory $\mathcal{M}=\{\mathcal{K}^{s},\mathcal{K}^{g},\mathcal{K}^{p}\}$, including their content, tags, and application conditions.

\definecolor{schemaBorder}{HTML}{DDE4E7}
\definecolor{schemaMuted}{HTML}{64737B}
\definecolor{schemaLilac}{HTML}{8176A0}
\definecolor{schemaTeal}{HTML}{5F8E8D}
\newcommand{\sampleschemaheading}[2][schemaMuted]{%
    {\sffamily\bfseries\fontsize{8.2}{10}\selectfont\textcolor{#1}{#2}\par}%
    \vspace{3pt}%
}
\newcommand{\sampleschemaentry}[9]{%
    \begin{tcolorbox}[
        enhanced, colback=#9!3!white, colframe=#9!28!white,
        boxrule=0.4pt, arc=4pt,
        borderline north={1.2pt}{0pt}{#9},
        left=11pt, right=11pt, top=12pt, bottom=12pt,
        before skip=0pt, after skip=0pt,
        fontupper=\fontsize{9.5}{12}\selectfont
    ]
    \begin{minipage}[t][258pt][t]{\linewidth}
        \RaggedRight\hyphenpenalty=10000\exhyphenpenalty=10000
        \tcbox[on line, colback=#9!13!white, colframe=#9!13!white,
            boxrule=0pt, arc=2pt, boxsep=0pt, left=5pt, right=5pt, top=2pt, bottom=2pt]
            {\sffamily\bfseries\fontsize{8.2}{10}\selectfont\textcolor{#9}{#3}}\hfill
        {\sffamily\bfseries\fontsize{8.2}{10}\selectfont\textcolor{#9}{#1}}\par
        \vspace{7pt}
        {\bfseries\fontsize{11}{13}\selectfont #2\par}
        \vspace{5pt}
        {\sffamily\fontsize{8.5}{10.5}\selectfont\color{schemaMuted}#4\enspace\textcolor{schemaBorder}{\textbullet}\enspace #5\par}
        \vspace{12pt}
        \begin{minipage}[t][54pt][t]{\linewidth}
            \sampleschemaheading{Description}
            #6\par
        \end{minipage}\par
        \vspace{10pt}
        \sampleschemaheading[#9]{Content}
        #7\par
        \vfill
        \begingroup
        \setlength{\fboxsep}{6pt}
        \noindent\colorbox{#9!8!white}{%
            \begin{minipage}{\dimexpr\linewidth-12pt\relax}
                \RaggedRight\hyphenpenalty=10000\exhyphenpenalty=10000
                \sampleschemaheading[#9]{When to use}
                {\fontsize{9}{11.5}\selectfont #8\par}
            \end{minipage}}
        \endgroup
    \end{minipage}
    \end{tcolorbox}%
}

\begin{table}[htbp]
    \centering
    \caption{Example schema for sample strategies ($K^s$) within the design memory $\mathcal{M}$.}
    \label{tab:sample_strategy}
    \vspace{4pt}
    \begin{minipage}[t]{0.485\linewidth}
        \sampleschemaentry{01}{ReAct Algorithm\\workflow trace}{Refiner}{History}{Flowchart}
            {Preserves a localized refinement record with source findings, applied edits, and post-edit diagnostics.}
            {Reasoning paths and feedback loops can be represented as explicit nodes and transitions when supported by the source content.}
            {When depicting reasoning steps, actions, and feedback loops in a flowchart.}{schemaLilac}
    \end{minipage}\hfill
    \begin{minipage}[t]{0.485\linewidth}
        \sampleschemaentry{02}{Microservices\\decomposition trace}{Structure}{CS}{Architecture}
            {Records the parsing of system descriptions into discrete service entities.}
            {High-entropy source text is filtered into a unique entity pool to reduce identity fragmentation in the diagram.}
            {When resolving topological complexity in system architecture graphs.}{schemaTeal}
    \end{minipage}
\end{table}

\newcommand{\knowledgecard}[9]{%
    \begin{tcolorbox}[
        enhanced, colback=#9!3!white, colframe=#9!28!white,
        boxrule=0.4pt, arc=4pt,
        borderline north={1.2pt}{0pt}{#9},
        left=12pt, right=12pt, top=11pt, bottom=11pt,
        before skip=0pt, after skip=0pt,
        fontupper=\fontsize{9.5}{12}\selectfont
    ]
        \RaggedRight\hyphenpenalty=10000\exhyphenpenalty=10000
        {\small
        \tcbox[on line, colback=#9!13!white, colframe=#9!13!white,
            boxrule=0pt, arc=2pt, boxsep=0pt, left=5pt, right=5pt, top=2pt, bottom=2pt]
            {\sffamily\bfseries\fontsize{8.2}{10}\selectfont\textcolor{#9}{#3}}}%
        \hspace{7pt}%
        {\sffamily\fontsize{8.5}{10.5}\selectfont\color{schemaMuted}%
            #4\enspace\textcolor{schemaBorder}{\textbullet}\enspace #5}%
        \hfill{\sffamily\bfseries\fontsize{8.2}{10}\selectfont\textcolor{#9}{#1}}\par
        \vspace{6pt}
        {\bfseries\fontsize{11}{13}\selectfont #2\par}
        \vspace{9pt}
        \sampleschemaheading{Description}
        #6\par
        \vspace{8pt}
        \sampleschemaheading[#9]{Content}
        #7\par
        \vspace{10pt}
        \begingroup
        \setlength{\fboxsep}{6pt}
        \noindent\colorbox{#9!8!white}{%
            \begin{minipage}{\dimexpr\linewidth-12pt\relax}
                \RaggedRight\hyphenpenalty=10000\exhyphenpenalty=10000
                \sampleschemaheading[#9]{When to use}
                {\fontsize{9}{11.5}\selectfont #8\par}
            \end{minipage}}
        \endgroup
    \end{tcolorbox}%
}

\begin{table}[!ht]
    \centering
    \caption{Example schema for domain guidelines ($K^g$) within the design memory $\mathcal{M}$.}
    \label{tab:domain guideline}
    \vspace{4pt}
    \knowledgecard{01}{Node icon positioning}{Layout}{History}{Flowchart}
        {Spatial placement rule for protecting text clarity.}
        {Place auxiliary icons outside dense text regions when the canvas has sufficient nearby whitespace.}
        {When adding icons to text-dense history flowcharts.}{schemaTeal}
    \vspace{10pt}
    \knowledgecard{02}{Time-series alignment}{Styling}{Finance}{Gantt}
        {Alignment rule for financial temporal data.}
        {Align time-series data horizontally and distinguish fiscal periods with consistent visual encodings.}
        {When displaying multiple fiscal periods in financial timelines.}{schemaLilac}
\end{table}

\begin{table}[!ht]
    \centering
    \caption{Example schema for general principles ($K^p$) within the design memory $\mathcal{M}$.}
    \label{tab:general_principle}
    \vspace{4pt}
    \knowledgecard{01}{Spatial exclusion for semantic clarity}{Layout}{Universal}{Universal}
        {Domain-agnostic rule distilled from supported lower-tier entries.}
        {Auxiliary elements should be positioned in available negative space adjacent to container boundaries to preserve information integrity.}
        {When a diagram contains optional annotations that may otherwise collide with core content.}{schemaTeal}
    \vspace{10pt}
    \knowledgecard{02}{Functional modularity}{Styling}{Universal}{Universal}
        {Cross-domain pattern emphasizing clear cluster separation.}
        {Visual hierarchy should map dominant attributes to critical entities while mapping peripheral nodes to muted styles.}
        {When establishing global consistency across chart types or styles.}{schemaLilac}
\end{table}

\clearpage
\section{End-to-End User Study}
\label{app:human-study}

\subsection{Human-Centered Web Application}

\begin{figure}[H]
    \centering
    \includegraphics[width=0.98\linewidth]{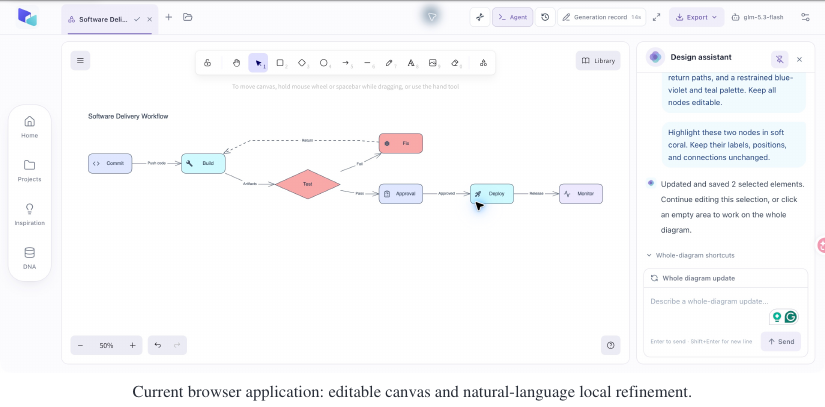}
    \caption{EvoDesign browser application. The workspace combines project and DNA navigation, an editable \texttt{excalidraw} canvas, export controls, and a design assistant. The conversation records a local request that updates two selected nodes while preserving the rest of the workflow.}
    \label{fig:system-demo}
    \vspace{-10pt}
\end{figure}

Figure~\ref{fig:system-demo} shows the project and design-memory navigation alongside the editable canvas and export controls.
The right-side design assistant records a selection-scoped request that updates two node fills without changing labels, positions, or connections.

\subsection{Human Study Setup}

We conducted a user study with 36 participants to evaluate EvoDesign in the two human-facing parts of the workflow that are not fully captured by automatic metrics: personalized design memory and editable refinement.
All participants had prior experience in diagram authoring.
They used the web application for their own diagramming tasks at least ten times within a day before completing the questionnaire.
The questionnaire asked about personalized memory, which learns design preferences from lightweight feedback, and editable refinement through the canvas schema.
Each item was rated on a five-point Likert scale from strongly disagree to strongly agree.

The study examined two groups of questions.
\textit{Personalized memory} asked whether participants perceived their own memory as better than generic memory, reported fewer edits, found the outputs readable, and considered their preferences captured.
\textit{Editable refinement} asked about perceived refinement speed, edit quality, workflow preference, and preservation of source meaning.
This setup follows the central use case of EvoDesign: generating an initial editable diagram, adapting it to user-specific design preferences, and supporting controlled refinement instead of forcing users to regenerate a static artifact.

Figure~\ref{fig:human-study-summary} in the main paper reports all eight item means and standard deviations.

\end{document}